\documentclass [fleqn,final,3p,times,number]{elsarticle}

\usepackage{graphicx,color,bm}
\usepackage{amsmath}
\usepackage{amssymb}

\usepackage{latexsym}
\usepackage{subfigure}
\usepackage{wrapfig} 
\usepackage[justification=justified, singlelinecheck=false]{caption}

\usepackage{multirow}
\usepackage{tabularx}
\usepackage[labelfont=bf]{caption}
\usepackage{rotating}
\usepackage{float,color}
\usepackage{caption}
\usepackage{subfig}
\usepackage[ampersand]{easylist}
\usepackage{colortbl}

\usepackage{tikz}
\usetikzlibrary{calc}
\floatstyle{plaintop}
\restylefloat{table}

\newcommand{\tarc}{\mbox{\large$\frown$}}
\newcommand{\arc}[1]{\stackrel{\tarc}{#1}}

\definecolor{Gray}{gray}{0.9}
\definecolor{grn1}{rgb}{0.0078,0.275,0.25}
\definecolor{blu1}{rgb}{0.0,0.0,0.5}

\journal{Journal of Computational Physics}

\begin{document}
\begin{frontmatter}
\title{Rotation matters - Direct numerical simulations of rectangular particles in suspensions at low to intermediate solid fraction}

\author[SU]{Zhipeng Qin\corref{cor1}}
\ead{zhipengq@stanford.edu}

\author[SU,UMD]{Kali Alison}
\ead{kallison@stanford.edu}

\author[SU]{Jenny Suckale} 
\ead{jsuckale@stanford.edu}

\address[SU]{Department of Geophysics, Stanford University, Stanford, California, USA}
\address[UMD]{Department of Geology, University of Maryland, College Park, Maryland, USA}

\cortext[cor1]{Corresponding author}

\begin{abstract}
Our ability to numerically model and understand the complex flow behavior of solid-bearing suspensions has increased significantly over the last couple of years, partly due to direct numerical simulations that compute flow around individual interfaces and hence resolve unprecedented detail. While most previous studies focus on spherical particles, we develop a direct numerical approach to capture rectangular particles. Our approach uses distributed Lagrange multipliers to enforce rigid-body motion in the solid domain in combination with an immersed boundary method to correctly enforce the no-slip constraint on the solid-fluid interfaces. An important component of our model is an efficient particle collision scheme that prevents overlap between particles of different shapes and allows for the transfer of both translational and angular momentum during particle collision. We verify and validate our numerical method through several benchmark cases. Applied to suspension flow, we test the hypothesize that particle rotations alter the aggregated dynamics of the suspension even if the relative rotational energy of the particles remains small as compared to the translational energy. At low solid fraction, we reproduce experimental observations of strongly nonlinear coupling between rectangular particles that is reminiscent of particle aggregation in the inertial regime but occurs at zero Reynolds number as a result of the long-range interaction between non-spherical particles. At intermediate solid fraction, we show that particle rotations can destabilize force chains. The dynamic consequences include the delayed onset of jamming and strong nonlinear coupling to the flow field in the fluid domain, which channelizes more strongly for rectangular as compared to spherical particles. While our model was motivated specifically by the need to better understand hazardous, crystal-bearing lava flows, our insights generalize to suspension flow in other scientific or engineering contexts. 
\end{abstract}
 
\end{frontmatter}

\section{Introduction}\label{sec:intro}
The behavior of small particles, drops, or bubbles in a viscous fluid at low Reynolds number is one of the oldest problems in theoretical fluid mechanics dating back as far as Stokes' analysis of a rigid sphere in an unbounded fluid from 1851 \citep{stokes1851effect}. Nonetheless, the flow dynamics of solid-bearing suspensions remain enigmatic \citep{brenner2001fluid}, despite decades of research progress as summarized in \citep{happel2012low}. One key challenge is that long-range interactions between particles lead to surprising spatial correlations in velocity fluctuations even at low solid fraction \citep{segre1997long, segre2001effective}. The resulting stochastic behavior is reminiscent of turbulence, even though the Reynolds number is very low \citep{xue1992diffusion, tong1998analogies, levine1998screened}. 

Despite the analogy between solid-bearing suspensions at low Reynolds number and turbulent flows, different modeling approaches are required for the two flows. In turbulent fluids, stochasticity emerges from dynamic instability in the fluid phase. Superimposing the motion of small, spherical particles on the fluid motion in a Lagrangian framework, which tracks particles but does not couple these back to the fluid phase, captures at least some of the important components of the aggregate dynamics \citep[e.g.,][]{falkovich2001particles, yeung2002lagrangian}. In contrast, stochasticity in solid-bearing suspensions at low Reynolds number emerges from the coupling between the solid and the fluid phase. A Lagrangian approach is hence questionable and a fully-coupled Eulerian model, where flow transports particles and particles reroute the flow around them and interact, is required. 

Direct numerical simulations of particulate flows solve the Stokes or Navier-Stokes equation at the scale of individual particles and offer a compelling, if computationally expensive, opportunity to compute the long-range hydrodynamic interactions without a-priori assumptions about drag forces or settling behavior. The accuracy of the numerical description, however, becomes delicate, which is not surprising for a highly nonlinear, multi-body boundary value problem. It is not sufficient to ensure rigid-body motion for solid particles \citep[e.g.,][]{Glowinski1999, Patankar2000}. It is equally important to enforce the no-slip boundary condition on the solid interfaces correctly \citep[e.g.,][]{Luo2007} and represent particle-particle collisions in a meaningful way \citep[e.g.,][]{Kempe2012, Costa2015, Biegert2017}.

To simplify the problem, most previous approaches have focused on spherical particles \citep[e.g.,][]{Glowinski1999, Patankar2000, Suckale2012a, breugem2012second}. While clearly a meaningful first step, analytical descriptions of the motion of single, non-spherical particles through singularity methods \citep{chwang1974hydromechanics, chwang1975hydromechanics, chwang1975hydromechanics3, chwang1976hydromechanics} or Euler angles \citep{gierszewski1978rotation, hinch1979rotation} have shown that non-spherical particles are more likely to rotate than spheres. The reason is that the area of the particle which is exposed to mean flow depends on the angle of rotation. As a consequence, the hydrodynamic drag force acting on the particle depends on the angle between particle and mean flow, as first noted by Becker \citep{becker1959effects}. Non-spherical shapes hence introduce an additional physical effect, which is largely absent in the spherical limit. 

Here, we hypothesize that the tendency of non-spheroidal particles to rotate affects suspension dynamics sensitively. To test this hypothesis, we develop a numerical methodology to better understand the dynamics of non-spherical particles in suspensions at low to intermediate solid fraction. We use direct numerical simulations to capture different crystal shapes building on \citep{Glowinski2001, ardekani2008collision, apte2009numerical}. The two main advances of our approach are (1) the usage of an immersed boundary method (IBM) following Qin and Suckale \citep{Qin2017} to improve solver accuracy in the vicinity of the solid-fluid interfaces and (2) the development of an efficient collision scheme between particles of arbitrary shape, which is necessary to be able to capture larger particle assemblages of non-spherical particles than Ardekani et al., \citep{ardekani2008collision}. 

The solid-fluid coupling we implement takes advantage of the fact that momentum is conserved in both the solid and fluid phase. Solving the Navier-Stokes equation in the entire domain hence assigns the correct momenta to the phases, but an additional constraint is needed to infer the velocity field in the solid phase from the computed momenta. Following Glowinski et al., \citep{Glowinski1999}, we use Lagrange multipliers distributed throughout the solid domain to project the fluid velocity field onto rigid body motion by enforcing that the deformation-rate tensor is zero there. Since the distributed Lagrange multipliers are only defined in the solid domain, they are not ideally suited for enforcing the correct boundary condition in the fluid adjacent to the solid boundary. 

To correct the velocity field in the fluid at the solid-liquid interface, we use an immersed boundary method. A number of immersed boundary methods have been developed in the past two decades as reviewed in Mittal and Iaccarino, \citep{Mittal2005}. In the present work, we implement a direct forcing strategy which is applicable to boundaries of arbitrary shape \citep[e.g.,][]{Luo2007} and is more efficient than other approaches such as the feedback forcing strategy \citep{Fadlun2000}. In our verification section, we take advantage of the analytical solution for Stokes flow around a circular cylinder in the vicinity of a moving wall to demonstrate that the proposed immersed boundary method affords higher order accuracy than previous direct forcing schemes and models that only rely on distributed Lagrange Multipliers to correct the solid-fluid boundaries \citep[e.g.,][]{ardekani2008collision, Suckale2012a}. 

\begin{figure}[t!]
	\centering
	\includegraphics[width=0.8\textwidth]{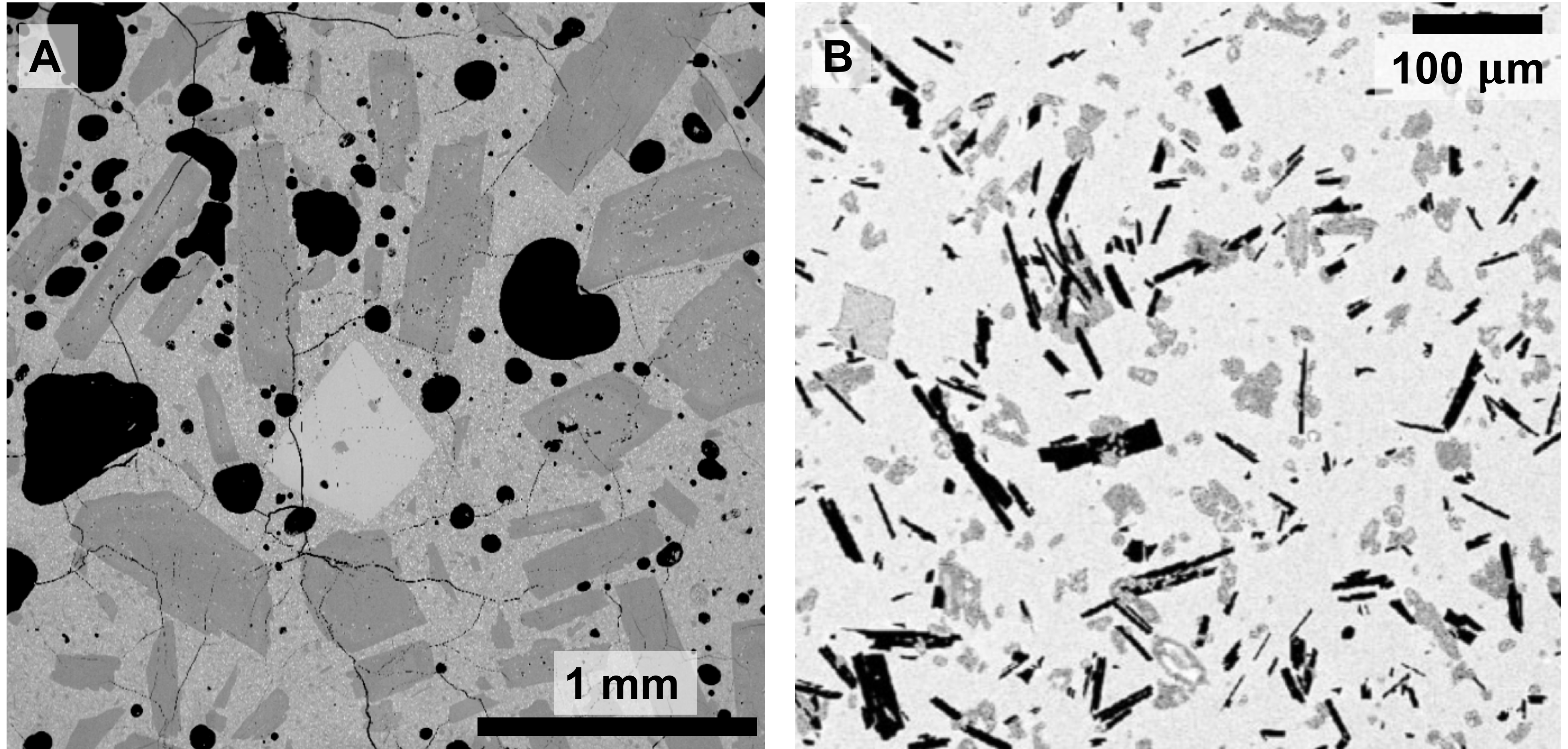}
	\caption{Examples of microtexture in lava flows. A: Two-dimensional section through a tephra sample erupted at Stromboli in July 2007 from \citep{belien2010gas}. Bubbles are shown in black, plagioclase crystals in dark grey, and pyroxene crystals in off-white. Light gray represents the formerly liquid portion of the sample, which underwent a glass transition upon erupting. B: Sample KE55-1888 from a subaerial lava flow at K${\mathrm{\bar{\imath}}}$lauea volcano, Hawaii, from  \citep{cashman1999cooling}. Plagioclase crystals are shown in black and pyroxene in dark grey. Figures courtesy of Katharine V. Cashman.}
	\label{fig:thinsections}
\end{figure}

To avoid unphysical overlap of the particles, we develop an efficient collision model for rectangular particles and mixed spherical and rectangular particles. Most previous studies of the behavior of non-spherical particles in suspensions assume spheroidal particles that have an axis of rotational symmetry \citep[e.g.,][]{jeffery1922motion, brenner1974rheology, petrie1999rheology}. We focus specifically on rectangular particles, because the development of this numerical technique was motivated specifically by the need to better understand hazardous lava flows. The minerals that form in lava flows during cooling typically have polygonal shapes (see Fig. \ref{fig:thinsections}). 

The dynamics of lava flows \citep[e.g.,][]{Hulme1974} and magma reservoirs \citep[e.g.,][]{mcintire2019hydrodynamics} depends on jamming phenomena induced by crystal clustering, which arise not only from the total crystal content but also from the long, elongated shape of some of the crystals such as plagioclase, which contribute a significant portion of the crystal load (e.g., Fig. \ref{fig:thinsections}A). Not surprisingly, existing laboratory work suggests that the presence of plagioclase crystals produces jamming at very low crystal fraction \citep[e.g.,][]{Philpotts1996, Philpotts1999}. The ability to resolve sharp, rectangular shapes is hence critical to advance our ability to model and understand lava flows.

Capturing clustering and jamming of suspended particles depends not only on resolving fluid flow around the solid-fluid boundary, but also requires a shape-adjusted collision or contact model. One commonly adopted collision scheme pioneered by \citep{Glowinski1999} applies a repulsive force shortly before particles come in contact. The repulsive potential is set up such that particles never come in direct physical contact or overlap. The scheme was improved \citep{Ardekani2008} by updating the particle velocity implicitly to avoid numerical instability and an overly restrictive Courant-Friedrichs-Lewy condition. 

While the repulsive-potential scheme is common in models of dilute flows \citep[e.g.,][]{uhlmann2008interface, lucci2010modulation, breugem2012second, santarelli2015direct, picano2015turbulent}, it becomes questionable in the limit of densely packed sediment beds, where particle-particle contact is ubiquitous \citep{Kempe2012, kempe2014relevance, Biegert2017}. The two main criticisms are that repulsive-potential methods introduce an artificial gap between particles and that they do not resolve the physics of collision and rebound, which is particularly important for understanding the mobilization of dense sediment beds \citep{nino1998experiments, lajeunesse2010bed}. 

In the limit of lava flows, the limitations are less consequential than for densely packed sediment beds. Microtextural analyses of crystal-bearing lava flows show that crystals often tend to be separated from each other by a finite melt film (e.g., Fig. \ref{fig:thinsections}A). Introducing a small but finite separation between particles into the numerical model is hence not unrealistic. In cases where this separation does not exist or is less clearly pronounced (e.g., Fig. \ref{fig:thinsections}B), the main mechanism contributing to clustering is that suspended crystals provide preferred nucleation sites, which leads to inter-growth of crystals. We neglect these geochemical effects here in the interest of simplicity. 

The physics of rebound also play a lesser role in magmatic flows than in sedimentary beds, because the shear viscosity of magmatic melts tends to be at least five orders of magnitude more viscous than the shear viscosity of water and often much higher than that. The impact Stokes number, St = $\rho_pv_0D/(9\mu_f)$, where $\rho_p$ represent the particle density, $D$ the particle diameter, $\mu_f$ the pure fluid viscosity, and $v_0$ the relative approach speed of the colliding particles, which governs rebound behavior, is hence typically much smaller than zero. For the validation cases that we consider later, St falls into the $10^{-6}$ to $10^{-7}$ range and can hence be neglected. Suppressed rebound is also consistent with the enduring contact of the particles observed in many field samples (e.g., Fig. \ref{fig:thinsections}). 

We hence argue that in the context of lava-flow modeling, the repulsive potential model proposed by Ardekani and Rangel \citep{Ardekani2008} provides a meaningful starting point. Here, we extend their scheme to particles with both spherical and rectangular shapes. To ensure that the repulsive forces prevent inter-penetration and satisfy the laws of Newtonian dynamics, we apply the analytical model proposed by Baraff \citep{Baraff1989} to formulate a linear system of inequality and equality constraints on the repulsive forces. The main advantage of this model is that it reduces the computation of the forces between systems of rigid bodies in collision to the solution of a linear programming problem which can be solved through a standard sparsity exploiting linear programming package, here we use the UMFPACK library written by Timothy Davis (\citep{Davis2006}). 



We have employed five benchmark cases to verify and validate our numerical model. To verify the numerical efficiency and mathematical accuracy of our solid-fluid coupling scheme, we take advantage of analytic solutions including the Wannier flow \citep{Wannier1950} and the flow field in a rotational viscometer \citep[e.g.,][]{Brady1988, Ishibashi2009}. To validate our solid-fluid coupling at finite Reynolds number, we compare our simulations to experimental results for flow past a square cylinder \citep{Dutta2009}. We then test whether our collision model is consistent with prior numerical results of spherical particles in a Couette device \citep{Karnis1966, Brady1988} and with experimental results of rectangular particles aggregating while sinking through a stagnant fluid \citep{Schwindinger1999}.  

\begin{figure}[h!]
	\centering
	\includegraphics[width=0.8\textwidth]{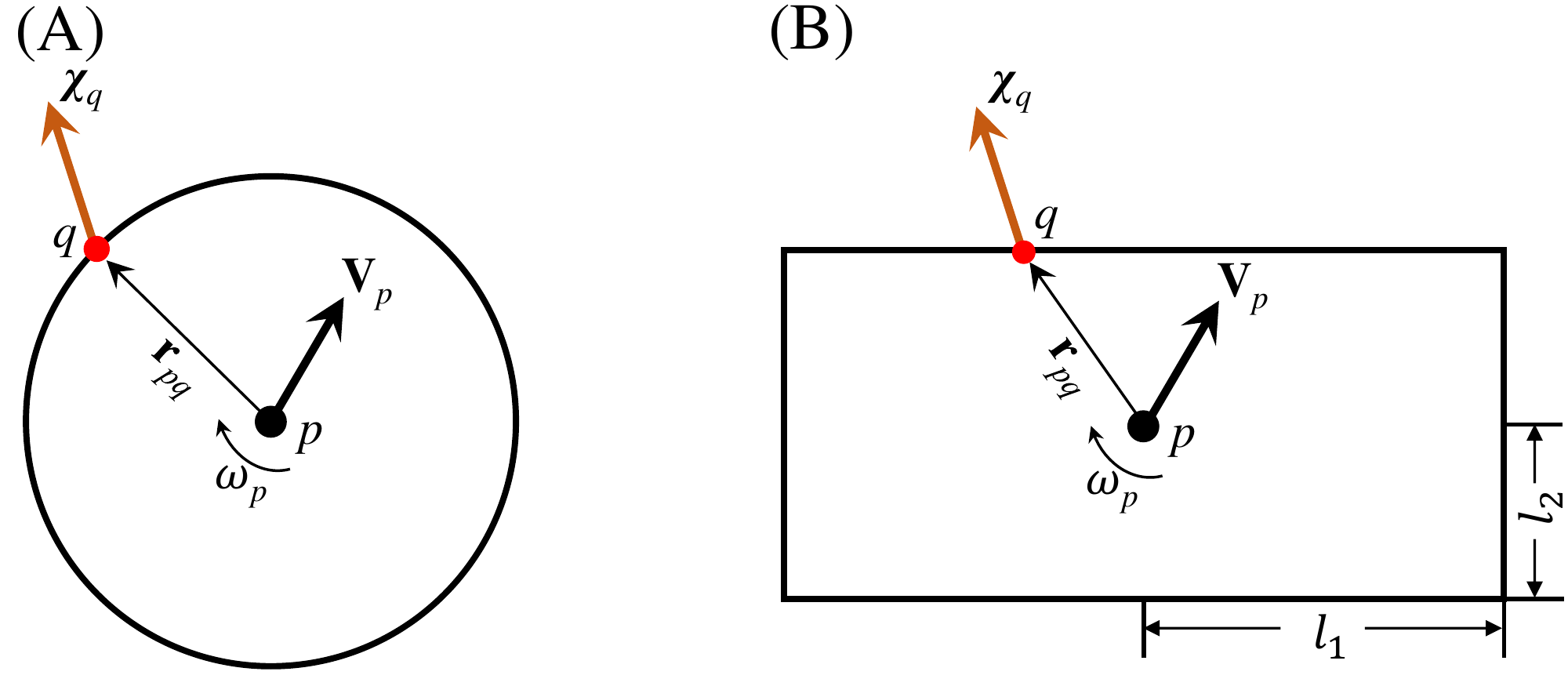}
	\caption{Diagram of single particles $p$ with two kinds of shapes, (A) circular; (B) rectangular. The linear velocity at the center of mass is ${\bf{V}}_p$, and the angular velocity is ${\bf{\omega}}_p$. At point $q$ on the edge, the velocity is ${\bf{\chi}}_q$.}
	\label{fig:singleRectangle}
\end{figure}

\section{Governing Equations} \label{sec:GE}
Our numerical model solves for conservation of mass and momentum in two dimensions. We assume that the pure liquid phase is incompressible. In the liquid domain, the governing equations are hence the incompressibility condition,
\begin{eqnarray} \label{eq:GE1}
	\nabla\cdot{\bf{v}}=0 \ ,
\end{eqnarray}
and the Stokes equation
\begin{eqnarray} \label{eq:GE2}
	-\frac{\nabla p}{\rho}+\frac{\mu}{\rho}\nabla^2{\bf{v}}+{\bf{g}}=0 \ ,
\end{eqnarray}
where $\bf{v}$ represents the velocity, $p$ the pressure, $\bf{g}$ the gravitational acceleration, $\mu$ the dynamic viscosity of the fluid, and $\rho$ the density of the fluid. We assume that the pure liquid phase abides by a Newtonian rheology. All non-Newtonian effects we detect are hence due to the interaction between solid and liquid phases. 

We represent the particles as rigid bodies fully immersed in the fluid. In the solid domain, the governing equations are hence Newton's equation of motion
\begin{eqnarray}\label{eq:GE3}
	\frac{M_pd{\bf{V}}_p}{dt}&=&{\bf{F}}_p+M_p\bf{g} \ , \\
	\label{eq:GE4}
	\frac{d({\bf{I}}_p{\bf{\omega}}_p)}{dt}&=&{\bf{T}}_p \ , \\
	\label{eq:GE5}
	\frac{d{\bf{X}}_p}{dt}&=&{\bf{V}}_p \ ,
\end{eqnarray}
where ${\bf{X}}_p$ is the position of the particles' center of mass, $M_p$ is the mass of the particle, ${\bf{V}}_p$ its velocity, ${\bf{\omega}}_p$ its angular velocity, ${\bf{I}}_p$ its angular moment of inertia tensor, and ${\bf{F}}_p$ and ${\bf{T}}_p$ represent the hydrodynamic force and torque exerted onto the particle by the surrounding fluid. 

\begin{figure}[h!]
	\centering
	\includegraphics[width=0.8\textwidth]{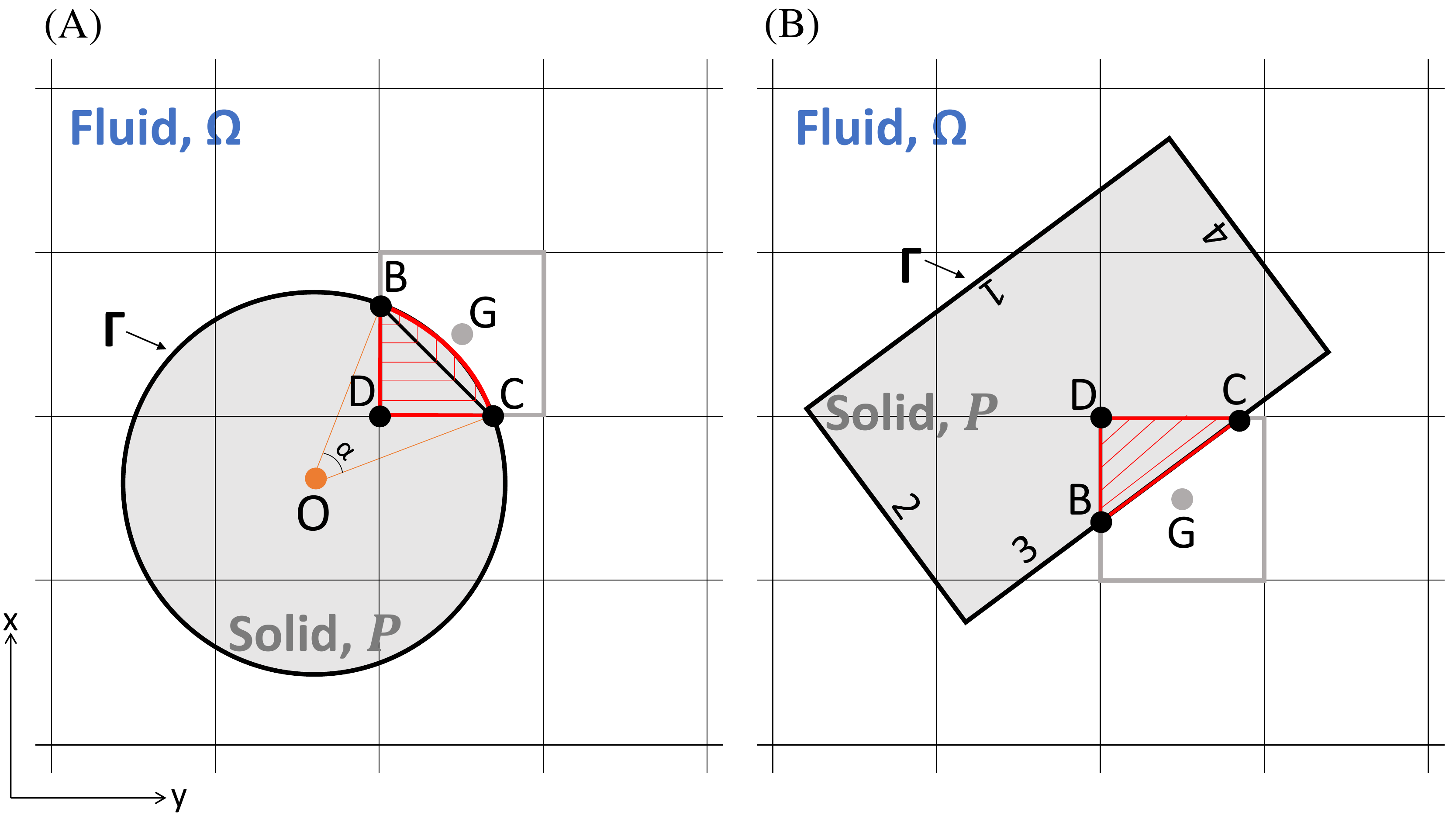}
	\caption{Illustration of the solid area in a computational cell. (A) represents the solid area in a computational cell for a circular disk; (B) represents the solid area in a computational cell a rectangular.}
	\label{fig:VF}
\end{figure}

In this paper, we focus specifically on circular and rectangular particles, primarily because of the importance of these shapes for lava flows (see Figure~\ref{fig:thinsections}). Simplified geometric forms as compared to general shapes are advantageous from a numerical point of view, because the moment of inertia tensor (see Figure~\ref{fig:singleRectangle}) and the solid fraction in a given computational cell (see Figure~\ref{fig:VF}) can be integrated analytically for idealized particle geometries. Some more details about these two analytical algorithms are introduced in following paragraphs. 

For a circular particle rotating around its center of mass, the angular moment of inertia tensor is ${\bf{I}}_p=M_p{\bf{r}}_{pq}^2$, where we define $||{\bf{r}}_{pq}||$ to be the radius of the circular particle (see Figure~\ref{fig:singleRectangle}A). For a rectangular particle (see Figure~\ref{fig:singleRectangle}B), ${\bf{I}}_p$ is
\begin{eqnarray} \label{eq:GE6}
	{\bf{I}}_p = ((2l_1)^2 + (2l_2)^2) M_p/12 \ .
\end{eqnarray}
While assuming idealized crystal shapes is advantageous because it significantly improves numerical accuracy, our numerical method lends itself to straight-forward extension to general particle shapes through a high-order integration scheme. 

In the interest of simplicity, we do not currently consider nucleation, growth, or dissolution of particles, implying that the velocity of point q on the edges of both circular and rectangular particles can be represented as
\begin{eqnarray}\label{eq:GE7}
{\bf{\chi}}_q={\bf{V}}_p+{\bf{\omega}}_p \times {\bf{r}}_{pq} \ .
\end{eqnarray}
This simplification implies the assumption that the time scale of crystal motion is faster than the time scale of crystal nucleation or growth, such that these process will only contribute a relatively negligible component to the overall dynamics of the flow. While clearly not generally true, we use this assumption as a first step towards a more complete understanding of the behavior of cyrstal-bearing magmatic flows.

In 2D, we define the solid fraction in a given computational cell as the ratio of the area occupied by particle to that of the whole cell. For the simplified geometric shapes we use, the solid fraction can be computed analytically by decomposing it into triangles, rectangles, and circular segments. Figure \ref{fig:VF} shows two examples for computing the solid fraction in cell, $G$. The first step in the analytical computation of the solid fraction is to identify how many grid vertexes fall into the solid domain. In Figure \ref{fig:VF}(A), only one vertex, $D$, is inside the solid domain and its coordinates satisfy the following condition,
\begin{eqnarray} \label{eq:60}
	-r + \sqrt{(x_D-x_O)^2+(y_D-y_O)^2} < 0 \ ,
\end{eqnarray}
where $r$ represents the radius of the circular disk, $x_D$ and $y_D$ are the x and y- coordinates of $D$, and $x_O$ and $y_O$ are the coordinates of the center of the disk, $O$. In the second step, we compute the coordinates of the intersection points between the interface and the cell faces, $B$ and $C$, through
\begin{eqnarray} \label{eq:61}
x_B = \sqrt{r^2 - (y_D-y_O)^2} + x_O \ ,
\end{eqnarray}
and,
\begin{eqnarray} \label{eq:62}
y_C = \sqrt{r^2 - (x_D-x_O)^2} + y_O \ ,
\end{eqnarray}
respectively. Finally, we calculate the solid fraction in cell $G$, $A_p$, as the sum of the triangle, $A_{BDC}$, and the circular sector cut off by the chord from B to C, $A_{\arc{BOC}}$,
\begin{eqnarray} \label{eq:63}
A_p = A_{BDC}+A_{\arc{BOC}}-A_{BOC}\ ,
\end{eqnarray}
where $A_{BDC}$($=\frac{1}{2}|BD||DC|$), $A_{BOC}$($=\frac{1}{2}|BO||CO|$) and
\begin{eqnarray} \label{eq:64}
A_{\arc{BOC}} = \frac{\alpha}{2\pi} \times 2\pi r\ ,
\end{eqnarray}
with $\alpha = \text{cos}^{-1}(\frac{OB \cdot OC}{|OB||OC|})$. 

Figure \ref{fig:VF}(B) illustrates an example for computing the solid fraction occupied by a rectangular particle. Similar to the computation for the circular disk, we first detect the number of vertexes inside the solid domain, $D$, by verifying
\begin{eqnarray} \label{eq:65}
\begin{split}
l_1^a x_D + l_1^b y_D + l_1^c < 0 \ , \\
l_2^a x_D + l_2^b y_D + l_2^c > 0 \ , \\
l_3^a x_D + l_3^b y_D + l_3^c > 0 \ , \\
l_4^a x_D + l_4^b y_D + l_4^c < 0 \ ,
\end{split}
\end{eqnarray}
where $l_{1-4}^a$, $l_{1-4}^b$ and $l_{1-4}^c$ are the coefficients of four line segments of the rectangle (marked in Figure \ref{fig:VF}(B)). We then compute $A_p$ as the area of the shaded triangle, $A_{BDC}=\frac{1}{2}|BD||DC|$, where the coordinate of the intersections $B$ and $C$ are,
\begin{eqnarray} \label{eq:66}
x_B = \frac{-l_3^b y_D - l_3^c}{l_3^a}  \ , 
\end{eqnarray} 
and,
\begin{eqnarray} \label{eq:67}
y_C = \frac{-l_3^a x_D - l_3^c}{l_3^b}  \ , 
\end{eqnarray} 
respectively. Needless to say, many more cases exist of how the grid might dissect the particles, but all follow the basic strategy of decomposing the solid fraction into simple geometric shapes for which the area can be computed analytically. An important advantage of this strategy is that the accuracy of the integration to obtain solid fraction does not depend on the orientation of the particle with respect to the grid, which can lead to numerical artifacts.

\section{Numerical Methodology} \label{sec:NM}
Due to the small size of the crystals (see Figure \ref{fig:thinsections}) and the high viscosity of the fluid, the Reynolds number is very small in many magmatic flows. As a consequence, the Navier-Stokes equation can often be reduced to the Stokes equation. While nominally easier because of the absence of the nonlinear term and the parabolic time dependence, the Stokes equation is a purely elliptic boundary-value problem. Approximating it numerically depends sensitively on the accuracy with which the numerous interior solid-fluid interfaces are represented. Here, we derive our fluid solver in the Stokes limit. At finite Reynolds number, we rely on the Navier-Stokes we have developed previously  \citep{Qin2017}.

The idea behind our methodology pioneered by Glowinski et al., \citep{Glowinski1999} is to solve the Stokes equation in the entire computational domain at the beginning of each time step. In the first step, both phases are hence treated as a fluid. In the second step, we use the distributed Lagrange multipliers to prevent deformation in the solid domain by projecting fluid onto solid motion, while keeping momentum conserved. In the third step, we enforce a no-slip boundary condition in the immediate vicinity of the solid-fluid interface, $\Gamma$, through our immersed boundary method. We hence generalize the one-phase momentum conservation (Eq. \ref{eq:GE2}) to the following two-phase equation,
\begin{eqnarray}\label{eq:NMN1}
	-\frac{\nabla p}{\rho(\bf{x})}+\frac{\mu}{\rho(\bf{x})}\nabla^2{\bf{v}}+{\bf{g}} + {\bf{f}}_{DLM}+{\bf{f}}_{IB}=0 \ ,
\end{eqnarray}
where, ${\bf{f}}_{DLM}$ is the body force solved by the distributed Lagrange multipliers, ${\bf{f}}_{IB}$ is the body force solved by the immersed boundary method. In the following subsections, we discuss the three main components of our numerical approach, namely 1) the projection-iterative solver of the Stokes equation on the Cartesian grid, 2) the distributed Lagrange multipliers; and 3) the immersed boundary method.

\subsection{Stokes solver}\label{sec:CG}
Our solver for the incompressible Stokes equation is based on the projection method \citep{Chorin1969} with approximate factorization \citep{Briley1977} and fractional time stepping \citep{Kim1985} on a staggered grid. Using this solver, we actually replace the governing equation (Eqs.~\ref{eq:GE1}-\ref{eq:GE2}) by the following equations,
\begin{eqnarray} \label{eq:NMN21}
\nabla\cdot{\bf{v}}=0 \ ,
\end{eqnarray}
\begin{eqnarray} \label{eq:NMN22}
\frac{\partial{\bf{v}}}{\partial t} = -\frac{\nabla p}{\rho}+\frac{\mu}{\rho}\nabla^2{\bf{v}}+{\bf{g}} \ ,
\end{eqnarray}
and solve them iteratively until the residual term on the left hand side falls below a predefined tolerance level, $\frac{\partial{\bf{v}}}{\partial t}<$ TOL$_S$. 

At each iteration n, we apply the projection method which treats the pressure term in Eq.~\ref{eq:NMN22} as a projection operator to project the initial velocity guess onto a divergence-free field (Eq.~\ref{eq:NMN21}). In the framework of the classical three projection steps, n, n+1/2 and n+1 \citep{Chorin1969}, we first obtained an provisional velocity field, $\hat{\bf{v}}$ at n+1/2,  
\begin{eqnarray}\label{eq:NMN2}
	\frac{\hat{\bf{v}}-{\bf{v}}^n}{\Delta t}=\frac{\mu}{\rho(\bf{x})}\nabla^2\hat{\bf{v}}+{\bf{g}}+{\bf{f}}_{IB}^{n} \,\,\,\,\,\,\,\, \text{in} \,\,\,\, \Omega \ ,
\end{eqnarray}
\begin{eqnarray}\label{eq:NMN3}
	\hat{\bf{v}} = {\bf{b}}^{n} \,\,\,\,\,\,\,\, \text{on} \,\,\,\, \Gamma \ ,
\end{eqnarray}
where ${\bf{f}}_{IB}^{n}$ and ${\bf{b}}^{n}$ are the boundary condition determined by the immersed boundary method. We treat the viscous term in Eq.~\ref{eq:NMN2} as an implicit term, which allows us to eliminate the time-step restriction associated with the viscous term. Hence, we rewrite Eq.~\ref{eq:NMN2} in two dimensions as
\begin{eqnarray}\label{eq:NMN4}
	\begin{split}
		\left[1-\frac{\mu\Delta t}{\rho({\bf{x}})}\frac{\partial^2}{\partial x^2}-\frac{\mu\Delta t}{\rho({\bf{x}})}\frac{\partial^2}{\partial y^2}\right] \hat{\bf{v}}={\bf{v}}^n+{\bf{g}}+{\bf{f}}_{IB}^{n} \ .
	\end{split} 
\end{eqnarray}
Replacing the large sparse coefficient matrix on the left hand side of Eq.~\ref{eq:NMN4} with tridiagonal matrices, we solve the following equation,
\begin{eqnarray}\label{eq:NMN5}
	\begin{split}
		&\left[1-\frac{\mu\Delta t}{\rho({\bf{x}})}\frac{\partial^2}{\partial x^2}\right]\left[1-\frac{\mu\Delta t}{\rho({\bf{x}})}\frac{\partial^2}{\partial y^2}\right] \hat{\bf{v}} ={\bf{v}}^n+{\bf{g}}+{\bf{f}}_{IB}^{n} \ ,
	\end{split} 
\end{eqnarray}
instead of Eq.~\ref{eq:NMN4}, associated with a second-order-accurate boundary condition \citep{Kim1985}, 
\begin{eqnarray}\label{eq:NMN6}
	\hat{\bf{v}}={\bf{b}}^{n}+\Delta t\frac{1}{\rho(\bf{x})}\nabla p^n \ . 
\end{eqnarray} 
Kim and Moin \citep{Kim1985} argued that, as an $O(\Delta t^3)$ approximation of Eq.~\ref{eq:NMN4}, Eq. \ref{eq:NMN5} results in a huge reduction in computing cost and memory. 

The provisional velocity field $\hat{\bf{v}}$ contributes to the final velocity field ${\bf{v}}^{n+1}$ and the gradient of the pressure later on,
\begin{eqnarray}\label{eq:NMN7}
	\frac{{\bf{v}}^{n+1}-\hat{\bf{v}}}{\Delta t}=-\frac{1}{\rho(\bf{x})}\nabla p^{n+1} \ ,
\end{eqnarray}
and
\begin{eqnarray} \label{eq:NMN8}
	\nabla\cdot{\bf{v}}^{n+1}=0 \,\,\,\,\,\,\,\, \text{in} \,\,\,\, \Omega \ ,
\end{eqnarray}
\begin{eqnarray}\label{eq:NMN9}
	{\bf{n}}\cdot{\bf{v}}^{n+1} = {\bf{n}}\cdot{\bf{b}}^{n+1} \,\,\,\,\,\,\,\, \text{on} \,\,\,\, \Gamma \ ,
\end{eqnarray}
where ${\bf{n}}$ represents the normal of the solid-fluid interface. 

We then solve the following Poisson equation
\begin{eqnarray}\label{eq:NMN10}
	\nabla\cdot(\frac{1}{\rho(\bf{x})}\nabla p^{n+1})=\frac{1}{\Delta t}\nabla\cdot\hat{\bf{v}}  \,\,\,\,\,\,\,\, \text{in} \,\,\,\, \Omega \ ,
\end{eqnarray}
implied by Eqs. \ref{eq:NMN7}-\ref{eq:NMN8} through an asymmetric multifrontal method \citep{Golub1989}. Introducing Eqs. \ref{eq:NMN3} and \ref{eq:NMN9} into Eq. \ref{eq:NMN7}, we obtain the following boundary condition for the pressure on a Dirichlet boundary,
\begin{eqnarray}\label{eq:NMN11}
	\nabla p^{n+1}\cdot{\bf{n}}=0 \,\,\,\,\,\,\,\, \text{on} \,\,\,\, \Gamma \ .
\end{eqnarray}
In the present work, we do not impose pressure boundary conditions on $\Gamma$ explicitly. As illustrated later, the momentum equation normal to $\Gamma$ reduces to d$p$/d$n$ = 0 on the boundary points, because we linearize all velocity components in the vicinity of $\Gamma$ in our immersed boundary method.

\subsection{Distributed Lagrange multipliers}\label{sec:DLM}

During the first iteration of the fluid solver, the additional body forces, ${\bf{f}}_{DLM}$ and ${\bf{f}}_{IB}$ are unknown.  We first solve the variable-density Navier-Stokes equation without these terms. To ensure that the deformation tensor in the solid domain is zero, 
\begin{eqnarray}\label{eq:NMN12-1}
	\boldsymbol{D}=\frac{1}{2}\left[\nabla{\bf{v}}+(\nabla{\bf{v}})^{\text{T}}\right]=0 \, ,
\end{eqnarray}
we use distributed Lagrange multipliers to project the initial fluid velocity field onto rigid-body motion \citep{Glowinski1999}. To estimate the linear and angular velocity of each particle, we take advantage of momentum conservation and integrate the fluid velocity inside the solid boundary to obtain an equivalent linear rigid-body velocity, 
\begin{eqnarray}\label{eq:NMN12-2}
	{\bf{V}}_{p}=\frac{1}{M_{p}}\int_{p}\rho_{p}{\bf{v}}^*dA\approx\frac{\rho_{p}}{M_{p}}\sum_{i,j}\Phi_{i,j}{\bf{v}}^*_{i,j} \ , 
\end{eqnarray}
and angular velocity,
\begin{eqnarray}\label{eq:NMN13}
	\begin{split}
		{\bf{\omega}}_{p}&=\frac{1}{{\bf{I}}_{p}}\int_{p}{\bf {r}}\times\rho_{p}{\bf{v}}^*dA\\
		&\approx\frac{1}{{\bf{I}}_{p}}\sum_{i,j}\Phi_{i,j}{\bf{r}}_{i,j}\times\rho_{p}{\bf{v}}^*_{i,j} \ .
	\end{split}
\end{eqnarray}
where ${\bf{v}}^*$ is the fluid velocity field solved from Stokes equation, which is represented by ${\bf{v}}^{n+1}$ in Section~\ref{sec:CG}, $\Phi_{i,j}=\frac{A_p}{\Delta x\Delta y}$ represents the solid volume fraction in cell $(i, j)$, and $A_p$ is the area inside the solid domain for a given computational cell, which is determined analytically by the scheme introduced in Section~\ref{sec:GE}. 

\begin{figure}[h!]
	\centering
	\includegraphics[width=0.8\textwidth]{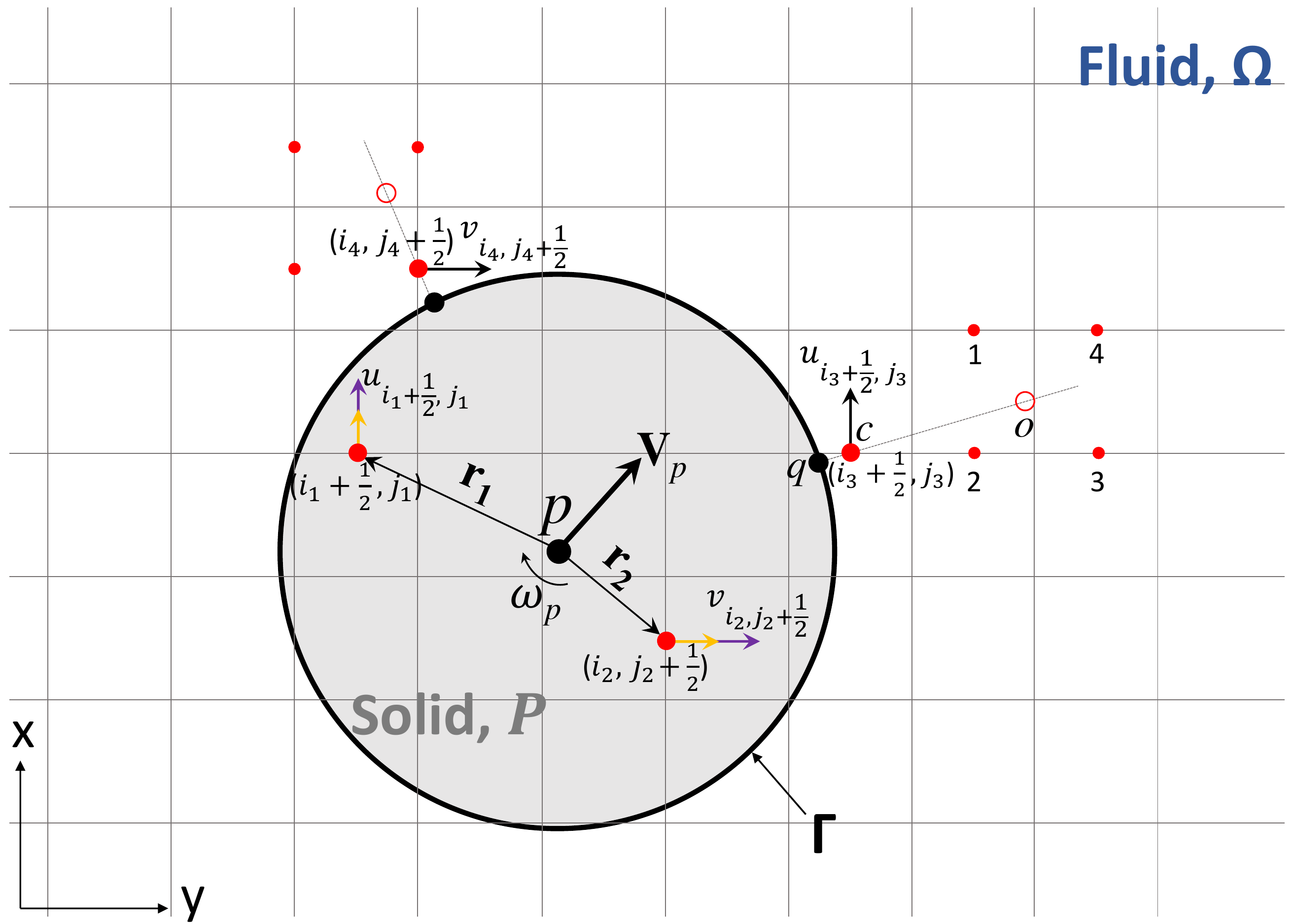}
	\caption{Illustration of our immersed boundary method. The velocities are stored on the staggered grids, for examples, u is on $(i_1+\frac{1}{2},j_1)$ and $(i_3+\frac{1}{2},j_3)$ while v is on $(i_2,j_2+\frac{1}{2})$ and $(i_4,j_4+\frac{1}{2})$. In the solid domain, the velocity is the sum of the linear component (yellow arrow) and the angular component (purple arrow). On the fluid in the immediate vicinity of the solid-fluid interface, we normally interpolate the velocity from the velocity on the interface, such as point $q$, and the velocity in the fluid domain further away from the interface, such as point $o$.}
	\label{fig:immersed}
\end{figure}

\subsection{Immersed boundary method}\label{sec:IBM}
One drawback of relying on the Navier-Stokes equation in both the fluid and solid domain is that the interface between the domains abides by a free-slip condition at least initially. Since the distributed Langrange multiplier are only defined in the solid domain, they may not correct the interface itself leading to an inaccurate representation of the no-slip condition on the solid-fluid boundary particularly in Stokes flow \citep[e.g.,][]{Qin2017}. To improve the representation of the no-slip condition, we use an immersed boundary method to derive the hydrodynamic forces that result from the no-slip condition in the immediate vicinity of the solid interface. 

Figure \ref{fig:immersed} illustrates our approach for the velocity in the solid domain and the velocity on the fluid in the immediate vicinity of the interface. Here, we introduce the x-component of this approach in detail, the y-component is similar. Since our solver is based on a staggered grid, $u$ is stored at the cell faces, such as $(i_1+\frac{1}{2},j_1)$ and $(i_3+\frac{1}{2},j_3)$ in Figure \ref{fig:immersed}. In the solid domain, we define the rigid body motion (see Eq. \ref{eq:GE7}) to obtain a corrected velocity, 
\begin{eqnarray}\label{eq:NMN14}
	u^{n+1}_{i_1+\frac{1}{2},j_1}=u_p-{\bf{\omega}}_p (y(j_1)-p_y) \ ,
\end{eqnarray}
where $p_y$ represents the y-coordinate of particle center, thus $(y(j_1)-p_y)$ represents the y-component of ${\bf{r}}_1$. 

In the fluid domain, e.g., at node c $(i_3+\frac{1}{2},j_3)$, we employ a direct forcing method to substitute $u^{*}_{i_3+\frac{1}{2},j_3}$, the fluid velocity solved from Stokes equation, with the weighted average velocity,
\begin{eqnarray}\label{eq:NMN151}
u^{n+1}_{i_3+\frac{1}{2},j_3}=\phi u_q+(1-\phi) u_f \ ,
\end{eqnarray}
where $\phi$ is an interpolating factor, $u_q$ is the solid velocity at point $q$ that is just on the solid-fluid interface and $u_f$ represents the fluid velocity without the influence of the solid-fluid interface. As demonstrated later, a linear interpolation scheme in the normal direction with respect to the interface as proposed by \citep{Balaras2004} enables a 2nd-order accurate representation of the boundary condition.

Other weighting approaches include Luo et al. \citep{Luo2007}, who proposed a nonlinear-weighted average scheme depending on the particle Reynolds number Re$_{p}=\frac{\rho_f|u_q-u_f|D}{\mu_f}$. In the framework of this scheme, $\phi$ is defined as follow,
\begin{eqnarray}\label{eq:NMN152}
	\phi = e^{-\text{Re}_p|X|} \ ,
\end{eqnarray}
where $|X|=\frac{|h|}{D}$ is the relative distance, $D$ is the crystal diameter, and $h$ is the distance from the node to the solid-fluid interface. In the Stokes regime, $\text{Re}_p \approx 0$, implying $\phi \approx 1$ and the solid velocity is imposed to the fluid in the immediate vicinity of the solid-fluid interface, which is not realistic. Another nonlinear strategy, which sets $u_f$ to  $u^{*}_{i_3+\frac{1}{2},j_3}$ and defines $\phi$ as the solid fraction ($\Phi$) in a computational cell, was also widely used in previous studies, but is limited to a 1st order accurate in space \citep{Fadlun2000}.

In terms of linear weightings, Qin and Suckale \citep{Qin2017} compute $u^{n+1}_{i_3+\frac{1}{2},j_3}$ through a 1D linear interpolation, defining $\phi = h_y/(h_y+\Delta y)$ and $u_f=u^{*}_{i_3+\frac{1}{2},j_3+\frac{1}{2}}$, where $h_y$ is the distance from the node to the interface in y-coordinate. While this scheme provides 2nd-order accuracy for spherical shapes, it is not suitable for complex shapes. For example, at the corners of the rectangular crystal considered here, it is ambiguous about which unique direction is that the interpolation can be performed over. The scheme proposed by Balaras \citep{Balaras2004} eliminates this ambiguity by linearly interpolating along the normal direction with respect to the interface. Figure \ref{fig:immersed} illustrates this scheme, by which $u_f$ is defined as the velocity at point $o$, $u_o$, and $\phi$ is defined as $\frac{|cq|}{|oq|}$. Note that the point $o$ is not located on a grid node, therefore it is needed to interpolate $u_o$ from the surrounding points by
\begin{eqnarray}\label{eq:NMN16}
	u_o=\sum^4_1\alpha_iu_i \ ,
\end{eqnarray} 
where $\alpha_i$ are the coefficients of a standard bilinear interpolation involving points 1-4. We hence follow the scheme by Balaras \citep{Balaras2004} here. 
\begin{figure}[h!]
	\centering
	\includegraphics[width=1\textwidth]{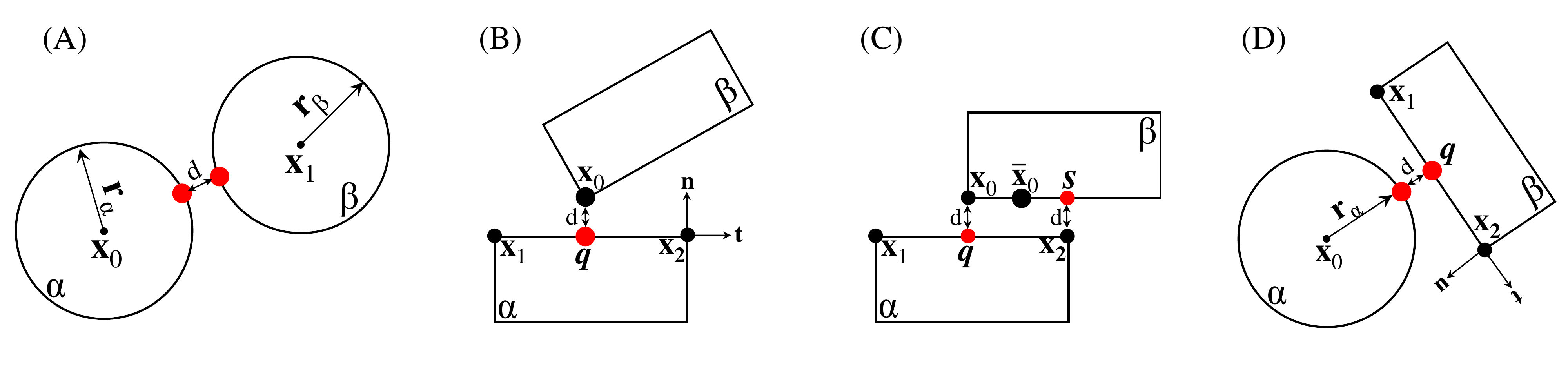}
	\caption{(a) Collision between two circular particles, the black solid circles are the center of particles and the red open circles are the closest points. (b) Collision between two rectangular particles, the black solid circles are the vertices of the rectangle and the red open circles are the closest points. (c) Collision between two parallel rectangular particles. (d) Collision between a rectangular particle and a circular particle, the black solid circles are the center of particles or the vertices of the rectangle, and the red open circles are the closest points. $d$ represents the distance between the particles $\alpha$ and $\beta$. $q$ and $s$ represent the points of potential collision. $\bf{t}$ is the normalized vector, and $\bf{n}$ is the vector perpendicular to $\bf{t}$.}
	\label{fig:detectingCollisionPoint}
\end{figure}

\section{Collision modeling}\label{sc:CM}
To model the collision and sustained contact between particles of different shapes, we implement a repulsive-potential model. More specifically, we build on the formulation by Ardekani and Rangel \citep{Ardekani2008} by updating the particle velocity implicitly to avoid numerical instability and an overly restrictive Courant-Friedrichs-Lewy condition. One consequence is that our particles never come directly in contact, which is consistent with field observations of lava flows showing a small but finite separation between crystals (see Figure \ref{fig:thinsections}). 

\subsection{Detecting a Collision Between Two Particles}
We define that a collision occurs when the closest distance between two particles, $d$, is less than or equal to a threshold distance $h$, where $h$ is selected to be twice of the grid spacing, $h=2\Delta x$, to ensure that the particles never overlap. Depending on the shapes of the colliding particles, we apply three different strategies to compute the closest distance $d$ and to identify the collision point (see Figure \ref{fig:detectingCollisionPoint}). 

When two circular particles collide, the collision point always lies on the line connecting the two centers (see Figure \ref{fig:detectingCollisionPoint}A). We compute the closest distance $d$ by subtracting the sum of two radius from the distance between two centers, such that $d = ||{\bf{x}}_1 - {\bf{x}}_0|| - (||{\bf{r}}_{\alpha}||+||{\bf{r}}_{\beta}||)$. The collision between two rectangles is more complex. We determine the closest distance $d$ by considering each particle as a collection of four points, the vertices of the rectangle, and four line segments, the edges of the rectangle. First, we compute the distance between each line segment of particle $\alpha$ and each vertex of particle $\beta$. The shortest distance over all of these computations can be denoted as $d_{\alpha\beta}$. Then, we compute the distance between each vertex of particle $\alpha$ and each line segment of particle $\beta$, and find the shortest distance over all of these computations, denoted as $d_{\beta\alpha}$. In the end, the distance $d$ is assigned to the shorter one of $d_{\alpha\beta}$ and $d_{\beta\alpha}$. 

The distance between a point ${\bf{x}}_0$  and a line segment defined by points ${\bf{x}}_1$  and ${\bf{x}}_2$, is computed based on whether the collision involves two vertices, a vertex and a edge (see Figure~\ref{fig:detectingCollisionPoint}B) or two edges (see Figure~\ref{fig:detectingCollisionPoint}C).

\begin{enumerate}
	\item {Collision between two vertices:} This is a special case of Figure~\ref{fig:detectingCollisionPoint}B. If $(\rm{min}(x_1,x_2) > x_0 \ \rm{or} \ \rm{max}(x_1,x_2) < x_0)$ and $ (\rm{min}(y_1,y_2) > y_0 \ \rm{or} \ \rm{max}(y_1,y_2) < y_0)$ where $x_{0, 1, 2}$ and $y_{0, 1, 2}$ are the components of ${\bf{x}}_{0, 1, 2}$, the distance is $\rm{min}(||{\bf{x}}_1-{\bf{x}}_0||,||{\bf{x}}_2-{\bf{x}}_0||)$.
	\item {Collision between a vertex and a edge:} If (1) is not the case, then let
	\begin{eqnarray}
		\begin{split}
			{\bf{t}} &= ({\bf{x}}_1-{\bf{x}}_2)/||{\bf{x}}_1-{\bf{x}}_2|| \nonumber \\
			{\bf{\hat{n}}} &= ({\bf{x}}_1-{\bf{x}}_0) - (({\bf{x}}_1-{\bf{x}}_0)\cdot {\bf{t}}) {\bf{t}}  \\
			q &= {\bf{x}}_0 + {\bf{\hat{n}}}  \nonumber \\
			d &= ||{\bf{\hat{n}}}|| \nonumber
		\end{split}
	\end{eqnarray}
	where ${\bf{t}}$ is a normalized vector pointing from ${\bf{x}}_1$ to ${\bf{x}}_2$, ${\bf{\hat{n}}}$ perpendicular to ${\bf{t}}$ is a vector whose length is the distance represented by $d$.
	\item {Collision between two edges:} The colliding particles are approximately parallel, as shown in Figure~\ref{fig:detectingCollisionPoint}C. We keep using above strategy but replace ${\bf{x}}_0$ by $\bar{{\bf{x}}}_0$ which the center point of ${\bf{x}}_0$ and $s$. 
\end{enumerate}

Finally, it is possible for a circular and a rectangular particle to collide, as illustrated in Figure \ref{fig:detectingCollisionPoint}D. This case arises in mixed shape simulations of when a circular particle collides with one of the domain walls. In this case, the line connecting the point of collision and the center of circular particle is perpendicular to the rectangular particle's closest edge to the center of the circular particle. The closest distance $d$ is then computed in a similar way to the one for two rectangles but using the radius of the circular object, $||{\bf{r}}_{\alpha}||$. 

\begin{enumerate}
	\item If $(\rm{min}(x_1,x_2) > x_0 \ \rm{or} \ \rm{max}(x_1,x_2) < x_0)$ or $ (\rm{min}(y_1,y_2) > y_0 \ \rm{or} \ \rm{max}(y_1,y_2) < y_0)$, then the distance is $\rm{min}(||{\bf{x}}_1-{\bf{x}}_0||,||{\bf{x}}_2-{\bf{x}}_0||)-||{\bf{r}}_{\alpha}||$.
	
	\item If (1) is not the case, then let
	\begin{eqnarray}
		\begin{split}
			{\bf{t}} &= ({\bf{x}}_1-{\bf{x}}_2)/||{\bf{x}}_1-{\bf{x}}_2|| \nonumber \\
			{\bf{\hat{n}}} &= ({\bf{x}}_1-{\bf{x}}_0) - (({\bf{x}}_1-{\bf{x}}_0)\cdot {\bf{t}}) {\bf{t}}  \\
			q &= {\bf{x}}_0 + {\bf{\hat{n}}}  \nonumber \\
			d &= ||{\bf{\hat{n}}}||-||{\bf{r}}_{\alpha}|| \nonumber
		\end{split}
	\end{eqnarray}
\end{enumerate}

\begin{figure}[h!]
	\centering
	\includegraphics[width=0.8\textwidth]{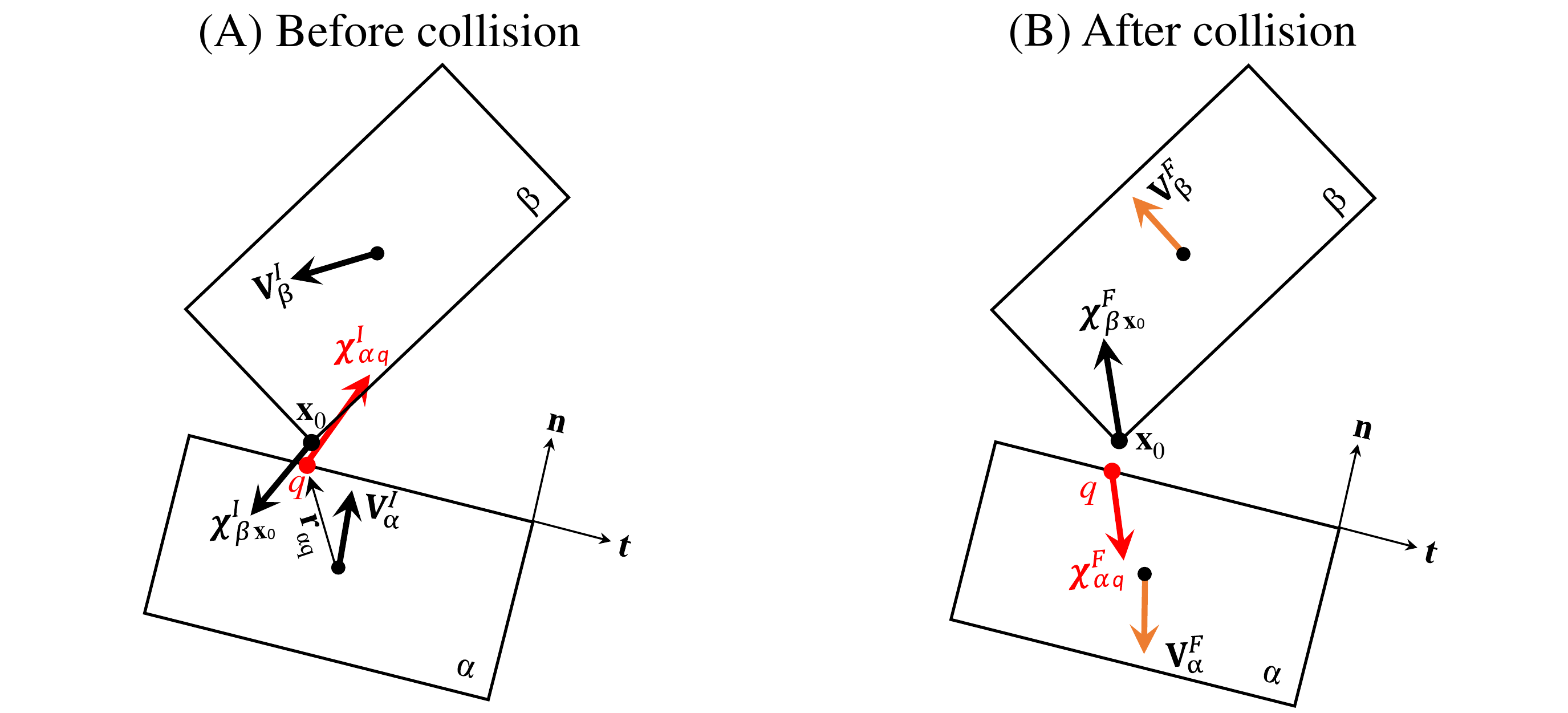}
	\caption{Velocities before (a) and after (b) the collision of two rectangular particles, $\alpha$ and $\beta$, at collision point $q$ and ${\bf{x}}_0$, respectively. These colliding particles are only restricted in normal direction but move freely in the tangential direction.}
	\label{fig:collision_2Rect}
\end{figure}

\subsection{Computing particle velocities after collision}
Suppose rectangles $\alpha$ and $\beta$ are colliding at points $q$ and ${\bf{x}}_0$, as illustrated in Figure~\ref{fig:collision_2Rect}. The initial linear velocities at each particles' center of mass are denoted with a superscript $I$ and subscript letter specifying the particle, such as ${\bf{V}}_{\alpha}^I$. The final central linear velocities, computed through our collision model, are denoted with a superscript $F$ and subscript letter specifying the particle, such as ${\bf{V}}_{\alpha}^F$. At the collision points, the velocities of each particle are 
\begin{eqnarray}
	\begin{split}
		{{\bf{\chi}}_{\alpha}}_q^{I(F)} &= {\bf{V}}_{\alpha}^{I(F)} + {\bf{\omega}}_{\alpha}^{I(F)} \times {\bf{r}}_{\alpha q} \\
		{{\bf{\chi}}_{\beta }}_{{\bf{x}}_0}^{I(F)} &= {\bf{V}}_{\beta}^{I(F)} + {\bf{\omega}}_{\beta}^{I(F)} \times {\bf{r}}_{\beta {\bf{x}}_0}
		\label{eq:finalVelocityA}
	\end{split}
\end{eqnarray}

Since there always is a thin liquid film between the colliding particles, the particles are free to move along the linear space spanned by the tangential vector, $\bf{t}$, without being affected by friction,
\begin{eqnarray}
	\begin{split}
		{{\bf{\chi}}_{\alpha}}_q^F\cdot {\bf{t}} &= {{\bf{\chi}}_{\alpha}}_q^I\cdot {\bf{t}} \\
		{{\bf{\chi}}_{\beta }}_{{\bf{x}}_0}^F\cdot {\bf{t}} &= {{\bf{\chi}}_{\beta}}_{{\bf{x}}_0}^I\cdot {\bf{t}}.
	\end{split}
	\label{eq:collisionConstraintA}
\end{eqnarray}
The velocities parallel to ${\bf{n}}$ are constrained by the conservation of linear momentum, such that
\begin{eqnarray}
	\begin{split}
		{{\bf{\chi}}_{\alpha}}_q^F\cdot {\bf{n}} - {{\bf{\chi}}_{\beta }}_{{\bf{x}}_0}^F \cdot ({\bf{n}}) = - c_R ({{\bf{\chi}}_{\alpha}}_q^I\cdot {\bf{n}} - {{\bf{\chi}}_{\beta }}_{{\bf{x}}_0}^I\cdot ({\bf{n}})),
		\label{eq:collisionConstraintB}
	\end{split}
\end{eqnarray}
where $c_R$ is the coefficient of restitution. For a purely inelastic collision, $c_R = 0$, and for a purely elastic collision $c_R = 1$. For all of the simulations in this paper, $c_R = 0$ since the viscosities are too high and the particles to small to be able to rebound. We compute the impact Stokes numbers for all validation cases involving multiple particles later to quantify this point better. 

Since the rectangles are rigid bodies, the change in the central velocity that occurs from the collision happens instantaneously and can be represented with a pair of impulses, $f{\bf{n}}$ and $f(-{\bf{n}})$,
\begin{eqnarray}
	\begin{split}
		{\bf{V}}_{\alpha}^F = {\bf{V}}_{\alpha}^I + \frac{f{\bf{n}}}{M_{\alpha}},\ \ \ & {\bf{\omega}}_{\alpha}^F = {\bf{\omega}}_{\alpha}^I + f \frac{{\bf{r}}_{\alpha q}\times {\bf{n}}}{I_{\alpha}}, \\
		{\bf{V}}_{\beta}^F = {\bf{V}}_{\beta}^I - \frac{f({\bf{n}})}{M_{\beta}},\ \ \ & {\bf{\omega}}_{\beta}^F = {\bf{\omega}}_{\beta}^I - f \frac{{\bf{r}}_{\beta {\bf{x}}_0}\times ({\bf{n}})}{I_{\beta}}.  
		\label{eq:finalVelocityB}
	\end{split}
\end{eqnarray}
Combining Eqs. \ref{eq:finalVelocityA} and \ref{eq:finalVelocityB} then giving them into Eq.~\ref{eq:collisionConstraintB}  produces an equation that can be solved for $f$
\begin{eqnarray}
	\begin{split}
		&\frac{{\bf{n}}\cdot{\bf{n}}}{M_{\alpha}} + \frac{{\bf{n}}\cdot{\bf{n}}}{M_{\beta}}
		+ \left(\frac{{\bf{r}}_{\alpha q}\times {\bf{n}}}{I_{\alpha}}\times{\bf{r}}_{\alpha q}\right)\cdot {{\bf{n}}}
		+ \left(\frac{{\bf{r}}_{\beta {\bf{x}}_0}\times {\bf{n}}}{I_{\beta}}\times{\bf{r}}_{\beta {\bf{x}}_0}\right)\cdot {\bf{n}} \\ 
		=& -\frac{1}{f}(1+c_R)({{\bf{\chi}}_{\alpha}}_q^I\cdot {\bf{n}} - {{\bf{\chi}}_{\beta}^I}_{{\bf{x}}_0}\cdot {\bf{n}})
		\label{eq:repulseforece}
	\end{split}
\end{eqnarray}

In the end, the corrected velocity of the colliding particles can be computed by giving $f$ and the constraint in the tangential direction, Eq.\ref{eq:collisionConstraintA}, into Eq.\ref{eq:finalVelocityB}.

\begin{figure}[h!]
	\centering
	\includegraphics[width=0.8\textwidth]{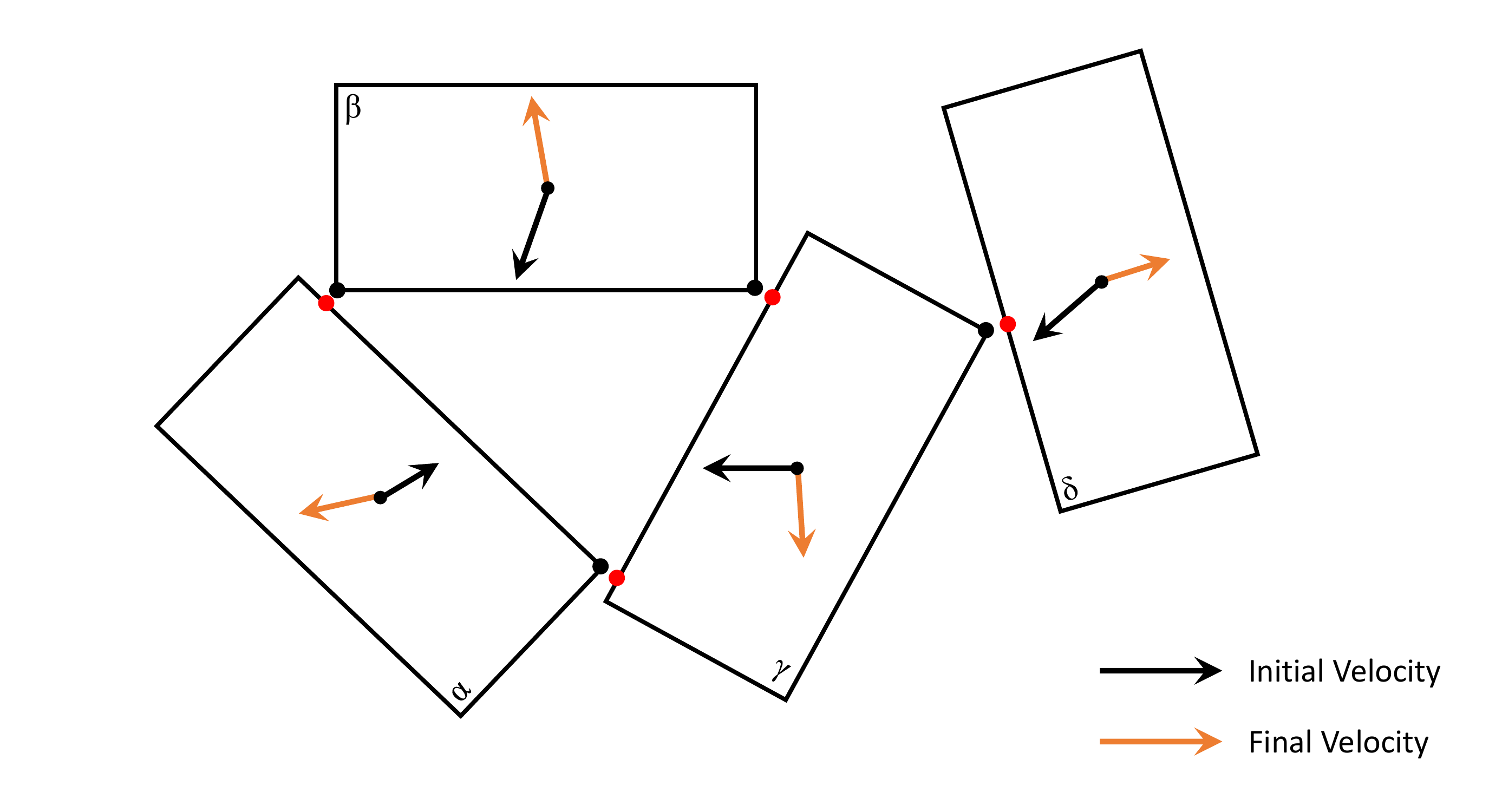}
	\caption{A system of four colliding rectangles, showing hypothetical initial velocities (black) and final velocities (yellow). In this case, $H=4$ and $N=4$, where $H>N-1$. }
	\label{fig:collision_3Rect}
\end{figure}

\subsection{Multiple-particle collisions}
As illustrated in Figure~\ref{fig:thinsections}, lava flows are characterized by long-lived clustering between multiple particles. To capture multiple-particle collisions, we generalize our collision model to capture force chain between systems of $N$ colliding particles, as shown in Figure \ref{fig:collision_3Rect}. In each cluster, one particle might collide with several other particles and hence experience a repulsive forces at the momentum of collision from multiple sources (e.g., particle $\alpha$ suffers the force from both $\beta$ and $\gamma$ in Figure \ref{fig:collision_3Rect}). Assuming the collision between the system of particles creates $H$ repulsive forces, we define the final linear and angular velocities at the center of a single particle through a generalized form of Eq. \ref{eq:finalVelocityB}, 
\begin{eqnarray}
	\begin{split}
		{\bf{V}}_{\xi}^F &= {\bf{V}}_{\xi}^I + \frac{1}{M_{\xi}}\sum_{\epsilon=1}^{H} \kappa f_{\epsilon}{\bf{n}}_{\xi\eta}, \ \ \xi = 1,...,N \\
		{\bf{\omega}}_{\xi}^F &= {\bf{\omega}}_{\xi}^I + \frac{1}{I_{\xi}} \sum_{\epsilon=1}^{H} \kappa f_{\epsilon}({\bf{r}}_{\xi q_{\xi\eta}} \times {\bf{n}}_{\xi\eta}) , \ \ \xi = 1,...,N
		\label{eq:multi-particle}
	\end{split}
\end{eqnarray}
where $\kappa=1$ when particle $\xi$ collides with particle $\eta$ while $\kappa=0$ when $\xi$ and $\eta$ do not collide, $f_{\epsilon}$ is the potential repulse force affecting between $\xi$ and $\eta$, ${\bf{n}}_{\xi\eta}$ is the normal vector from $\xi$ to $\eta$, ${\bf{r}}_{\xi q_{\xi\eta}}$ is the vector from the center of $\xi$ to the collision point, $q_{\xi\eta}$. Note that $q_{\xi\eta} \neq q_{\eta\xi}$ because of the thin separating film introduced in above sections, like the black points versus the red points shown in Figure \ref{fig:collision_3Rect}.

Applying above definition of the final velocities of each single particle (Eq. \ref{eq:multi-particle}) into the non-overlapping constraint (similar to Eq. \ref{eq:collisionConstraintB}) for each colliding pairs, we formulate a $H \times H$ linear system to compute the net forces as follow,
\begin{eqnarray}
	{\bf{A}} {\bf{f}} = {\bf{b}} \ ,
	\label{eq:linearsystem}
\end{eqnarray}
which can be seen as an extension of Eq. \ref{eq:repulseforece}. In above linear system, $\bf{A}$ represents the coefficient matrix, where the rows corresponds with the particles involved in each collision pair between particles $k$ and $l$, and the columns corresponds with the repulse forces affecting on the particles $i$ and $j$, so
\begin{eqnarray}
	\mathbf{A} =
	\begin{bmatrix}
		a^{12}_{12} & a^{12}_{13} & \dots  & a^{12}_{ij} \\
		a^{13}_{12} & a^{13}_{13} & \dots  & a^{13}_{ij} \\
		\vdots      & \vdots      & \ddots & \vdots      \\
		a^{kl}_{12} & a^{kl}_{13} & \dots  & a^{kl}_{ij}
	\end{bmatrix}
\end{eqnarray}
where $i < j$, $k < l$. Each element in $\mathbf{A}$ is given by

\begin{eqnarray}
	a^{kl}_{ij} =
	\begin{cases}
		\alpha^{kl}_{kl} + \alpha^{lk}_{lk} \ ,     & \small{\text{if } i = k, j = l } \\
		\alpha^{kl}_{kj} \ ,                        & \small{\text{if } i = k, j \neq l } \\
		\alpha^{kl}_{ki} \ ,                        & \small{\text{if } j = k, i \neq l } \\
		\alpha^{kl}_{lj} \ ,                        & \small{\text{if } j \neq k, i = l} \\
		\alpha^{kl}_{li} \ ,                        & \small{\text{if } i \neq k, j = l} \\
		0                \ ,                        & \small{\text{if } k \neq i, l \neq j} \\
	\end{cases}
\end{eqnarray}
where
\begin{eqnarray}
	\alpha_{ij}^{kl} = \frac{{\bf{n}}_{ij}\cdot{\bf{n}}_{kl}}{M_i} + \frac{{\bf{r}}_{iq_{ij}} \times {\bf{n}}_{ij}}{I_i}\times {\bf{r}}_{iq_{kl}}\cdot{\bf{n}}_{kl} \ .
\end{eqnarray}

We solve Eq. \ref{eq:linearsystem} by a sparsity exploiting linear programming package, UMFPACK (Unsymmetric MultiFrontal method and direct sparse LU factorization). 

\begin{figure}[h!]
	\centering
	\includegraphics[width=1\textwidth]{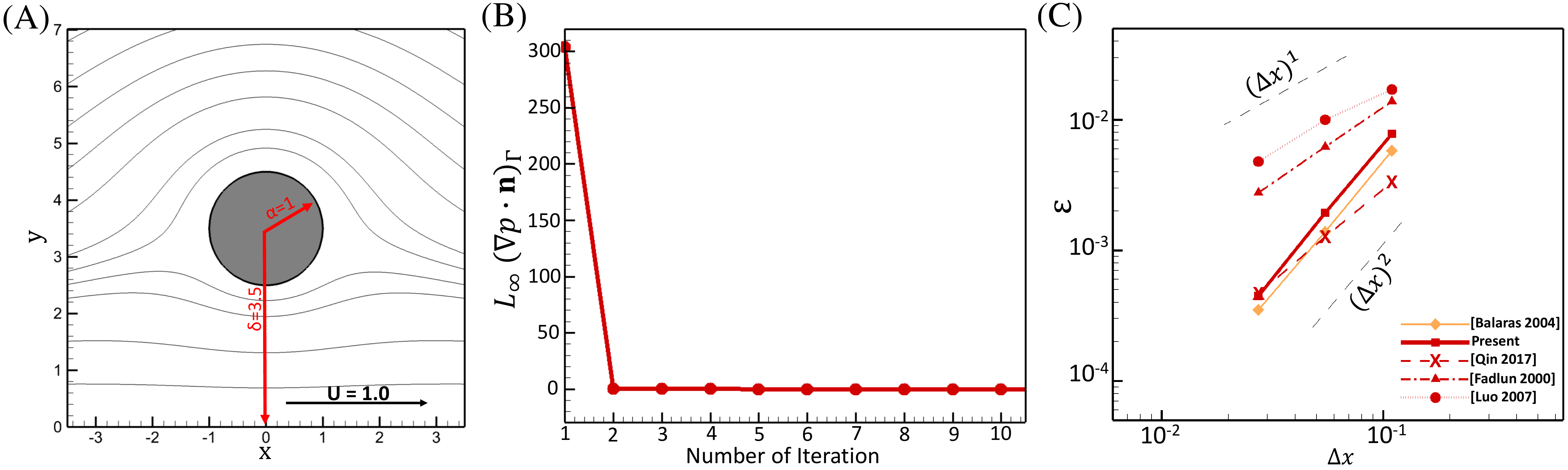}
	\caption{Wannier flow test case. (A) Computational domain and computed streamlines; (B) Evolution of the maximum derivative of the pressure along the normal direction of the solid-liquid interface on the grid of 128$\times$128; (C) Convergence study of the error for velocity in x-components, several schemes are compared here, yellow line represents the result given by Balaras \citep{Balaras2004}; red solid line represents the result given by the present numerical model using the normally linear interpolation; red dashed line is the result given by method from \citep{Qin2017}; the red dot dashed line is the result given by the method from \citep{Fadlun2000}; the red dotted line is the result given by the method using nonlinear-weighted average scheme \citep{Luo2007}.}
	\label{fig:Wannier}
\end{figure}

\section{Verification}\label{sec:VF}
We are not aware of any analytical solutions of flow around rectangular bodies that we can use for verification purposes. We hence verify the accuracy of our coupled fluid-solid solver for two circular geometries.  In the first case, we verify our numerical technique by reproducing the case of Stokes flow around a fixed cylinder in the vicinity of a moving plate. In the second case, we compute the velocity profile in a rotational viscometer to test whether our model correctly enforces the no-slip boundary condition on the moving solid-fluid interfaces. 

\subsection{Wannier flow} 

Stokes flow around a circular cylinder in the vicinity of a moving wall is named Wannier flow honoring an analytical solution by Wannier \citep{Wannier1950}. Using the analytical solution, we can quantify the numerical error of our coupled solid-fluid solver. Figure \ref{fig:Wannier}(A) shows the flow configuration and the computed streamlines on the grid of 64$\times$64. The Reynolds number for this case is $\sim$10$^{-3}$. In the computation, we assign a moving wall boundary condition to the bottom boundary and the analytical solution on all other boundaries. 

To test whether our simulation method correctly enforces the no-slip condition, we plot evolution of the maximum normal derivative of the pressure on the solid-liquid interface in the infinity norm (see Figure \ref{fig:Wannier}B). The normal derivative of the pressure reduces to 1\% of the initial value during the second iteration and decreases to less than 10$^{-4}$  after a few iterations. We can hence capture the no-slip boundary condition iteratively, although we did not explicitly impose a pressure boundary condition.

Figure \ref{fig:Wannier}(C) shows the $L_2$ norm of the error between the numerical and the analytical solution of the velocity components in x direction. We implement and compare several immersed boundary methods here to compare their convergence properties. We find that the linear-weighting scheme by Balaras \citep{Balaras2004} and our implementation afford second-order convergence rates while the nonlinear-weighting schemes \citep{Fadlun2000, Luo2007} are limited to approximately first order. 

\begin{figure}[h!]
	\centering
	\includegraphics[width=1\textwidth]{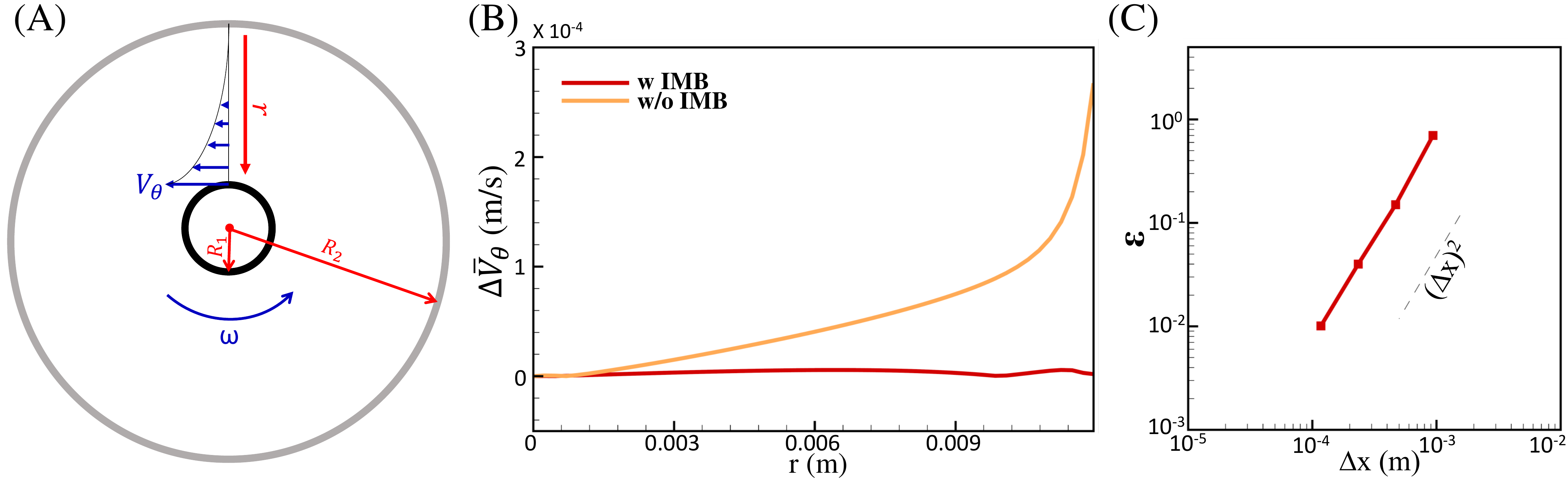}
	\caption{(A) The 2D model of a rotational viscometer. (B) Difference between the numerical radial velocity $V_{\theta}$ averaged over all $\theta$ (red) and the theoretical velocity in the radial direction. As a comparison, the numerical result without applying immersed boundary method is plot as yellow curve. (C) A convergence study for the $L_{\infty}$ norm of the error between numerical and theoretical solution.}
	\label{fig:rotation}
\end{figure}


\subsection{Rotational viscometer}\label{sec:RV}
Rotational viscometers are routinely used in laboratory studies to quantify the rheological properties of solid-bearing suspensions \citep[e.g.,][]{Ishibashi2009, mueller2009rheology}. Similar to Wannier flow, the velocity profile is sensitively dependent on enforcing the no-slip condition correctly. We hence use this case to demonstrate the value of the immersed boundary method. Figure \ref{fig:rotation}(A) shows our 2D representation of a rotational viscometer that consists of a fixed outer cylinder (gray circle) and an inner cylinder (black line) rotating at constant angular velocity, $\omega$. We set the radii of the outer and inner cylinder to be same as the rotational viscometer used in \citep{Ishibashi2009}, where $R_1$ = 0.003 m and $R_2$ = 0.015 m. For Stokes flow of a Newtonian fluid, the velocity profile, $V_{\theta}$, can be solved for analytically as,
\begin{eqnarray} \label{eq:ver1}
V_{\theta}=\frac{\omega}{R_2^2-R_1^2}(-R_1^2r+\frac{(R_1R_2)^2}{r}) \ ,
\end{eqnarray}
where $r$ represents the radial coordinate. 

In the numerical simulation, we treat the cross sections of both cylinders as circular disks and enforce a finite angular velocity, $\omega=1$, on the inner cylinder and a zero angular velocity on the outer cylinder. The velocity of the center is zero for both disks. In the gap between two disks, we set the density and the viscosity of the fluid to, $\rho_f=1000$ kg/m$^3$ and $\mu_f=1000$ Pa s, respectively. Because of the small length scale and the high viscosity, the Reynolds number is approximately $10^{-5}$. It is hence reasonable to consider the fluid as a Stokes fluid. 

Figure \ref{fig:rotation} (B) shows the difference between the numerical radial velocity averaged over all $\theta$ on a grid of 128$\times$128 and the theoretical solution given by Eq. \ref{eq:ver1}. We also plot the numerical result based only on distributed Lagrange Multipliers without enforcing the immersed boundary method and observe a significant drop in numerical accuracy. The drop of the numerical error quantified through the $L_{\infty}$ shows that we obtain second order convergence (see Figure \ref{fig:rotation} C).

\begin{figure}[h!]
	\centering
	\includegraphics[width=0.8\textwidth]{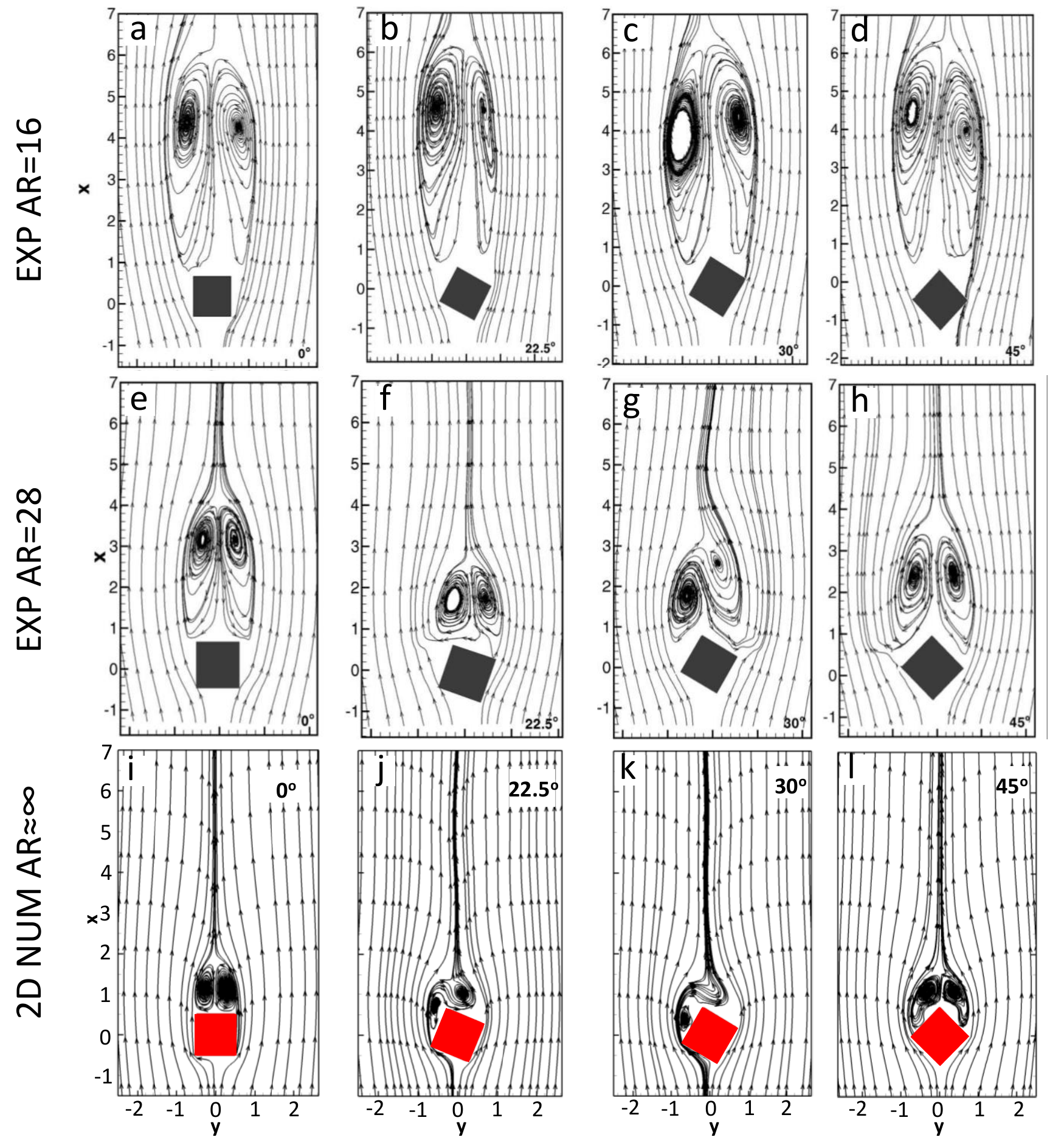}
	\caption{Comparison between the stream lines from experiment and numerical simulation at $\mathrm{Re} = 410$. x- and y- coordinates are expressed in units of the edge of the square. (a)-(d) represent the time-average stream trances in the wake of a square cylinder whose AR is 16; (e)-(h) represent the time-average stream trances in the wake of a square cylinder whose AR is 28; (i)-(l) represent the time-average stream lines in the wake of a square given by our 2D numerical model.}
	\label{fig:streamline}
\end{figure}

\section{Validation}\label{sec:VD}
Over the last couple of years, an increasing number of experimental studies have considered rectangular shapes that we can use for validation purposes. Here, we investigate flow past a rectangular cylinder at intermediate Reynolds number \citep{Dutta2009}, joint settling of rectangular particles in a stagnant fluid \citep{Schwindinger1999} and jamming of spherical particles in a shear suspension \citep{Brady1988} as test cases. 

\subsection{Flow past a square cylinder}
To assess that our numerical scheme correctly resolves the impact of geometry on flow in the immediate vicinity of the boundary, we simulate the two-dimensional flow over a fixed square cylinder at $\mathrm{Re} = 410$ and compare our numerical results to the analogue experiments by Dutta et al., \citep{Dutta2009}. Similar to the experimental setup, we fix a square cylinder of 3.4 mm edge (labeled "D") on the centerline of the computational domain whose width, W, is 4.8 cm. The boundary conditions imposed on the flow field are no-slip, ${\bf{v}}=0$, along the sidewalls. The inflow boundary is constant velocity, $u_{\text{inf}}$, and the outflow boundary, defined on the opposite side, is solved by a convective flow condition \citep{Orlanski1976},
\begin{eqnarray}\label{eq:Bech6}
	\frac{\partial{\bf{v}}}{\partial t}+u_{\text{inf}}\frac{\partial{\bf{v}}}{\partial {\bf{n}}}=0 \ , 
\end{eqnarray}
where $\bf{n}$ is the normal vector on the boundary. 

\begin{figure}[h!]
	\centering
	\includegraphics[width=1\textwidth]{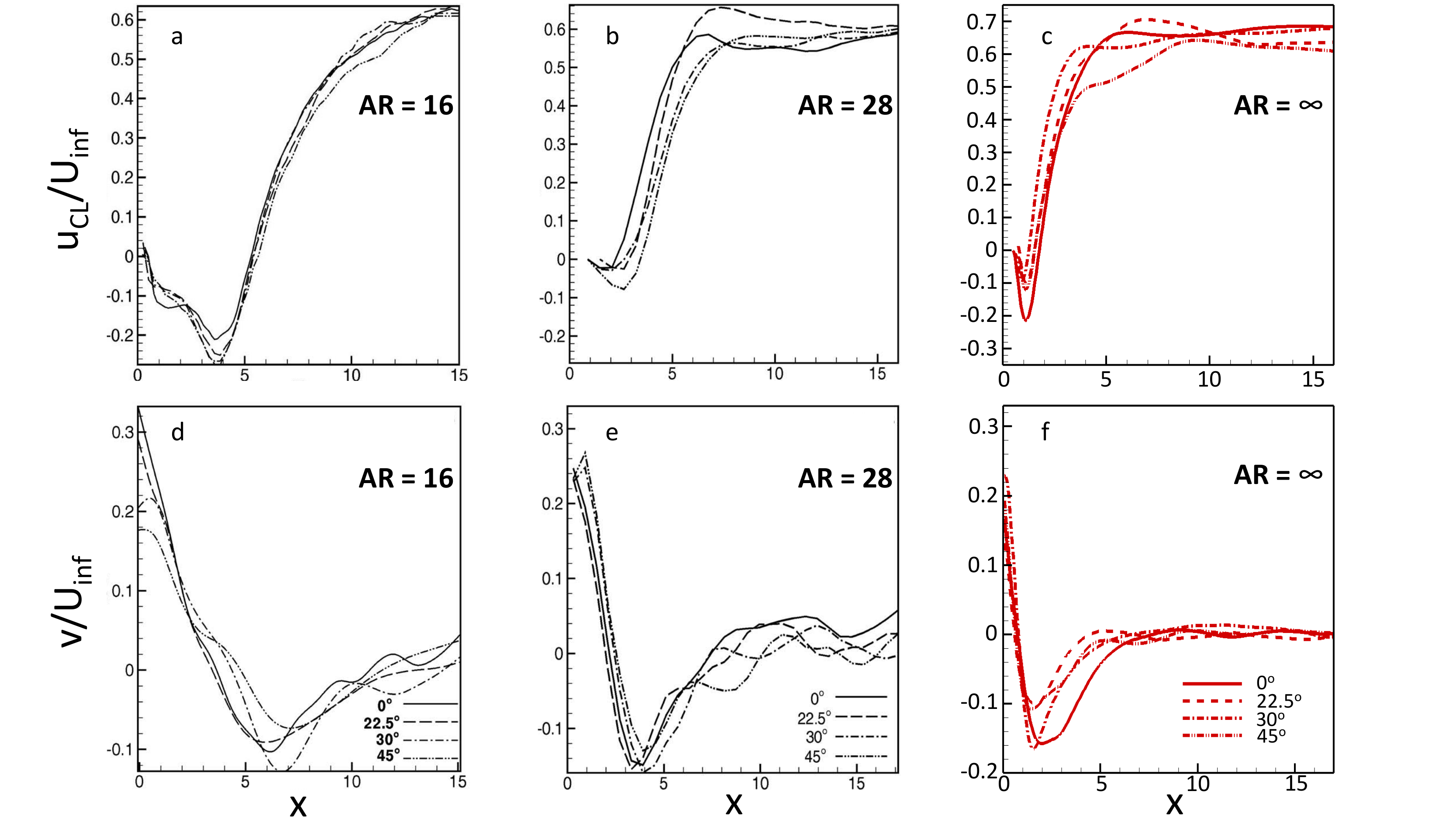}
	\caption{Centerline recovery of streamwise (u) and transverse (v) velocity component for four cylinder orientations at AR = 16 (a and d), AR = 28 (b and e) and a 2D square in the numerical model (c and f). x-coordinates is expressed in units of the edge of the square, and velocities are expressed in units of the inflow.}
	\label{fig:centerline}
\end{figure}

Figure \ref{fig:streamline} compares the contours of the numerical streamfuction to the experimental streak lines for four different cylinder orientation ($\theta=0, 22.5, 30, 45^\circ$), and aspect ratios (16 and 28 for experiments and $\infty$ for the numerical simulation), where the aspect ratio is defined as the ratio of the height of the cylinder to the edge of its cross-section. The size of the corresponding computational domain is W$\times$2W, and the simulation is based on a resolution of $200\times400$. In both the numerical simulation and the experimental results, the flow around the cylinder separates from the surface of the square and forms two recirculation vortex in the near wake. The unequal vortices presented in panels b, c, f, g, i and j demonstrate that the flow is asymmetric when the cylinder orientation is 22.5 or $30^\circ$. Comparing the width and length of the vortex between the cases with AR = 16 and AR = 28, we find that the vortices shrink at higher aspect ratio for all cylinder orientations. Moreover, the distance between the vortex and the square decreases at higher aspect ratio. These differences between higher and lower aspect ratio are likely a consequence of the three dimensionality of the flow field \citep{Dutta2009}. The 2D numerical simulation, which by its lower dimensionality implicitly assumes that the aspect ratio is infinite large, shows similar trends. Comparing panels i-l to e-h, we observe a smaller vortex that is much closer to the square.

\begin{figure}[h!]
	\centering
	\includegraphics[width=1\textwidth]{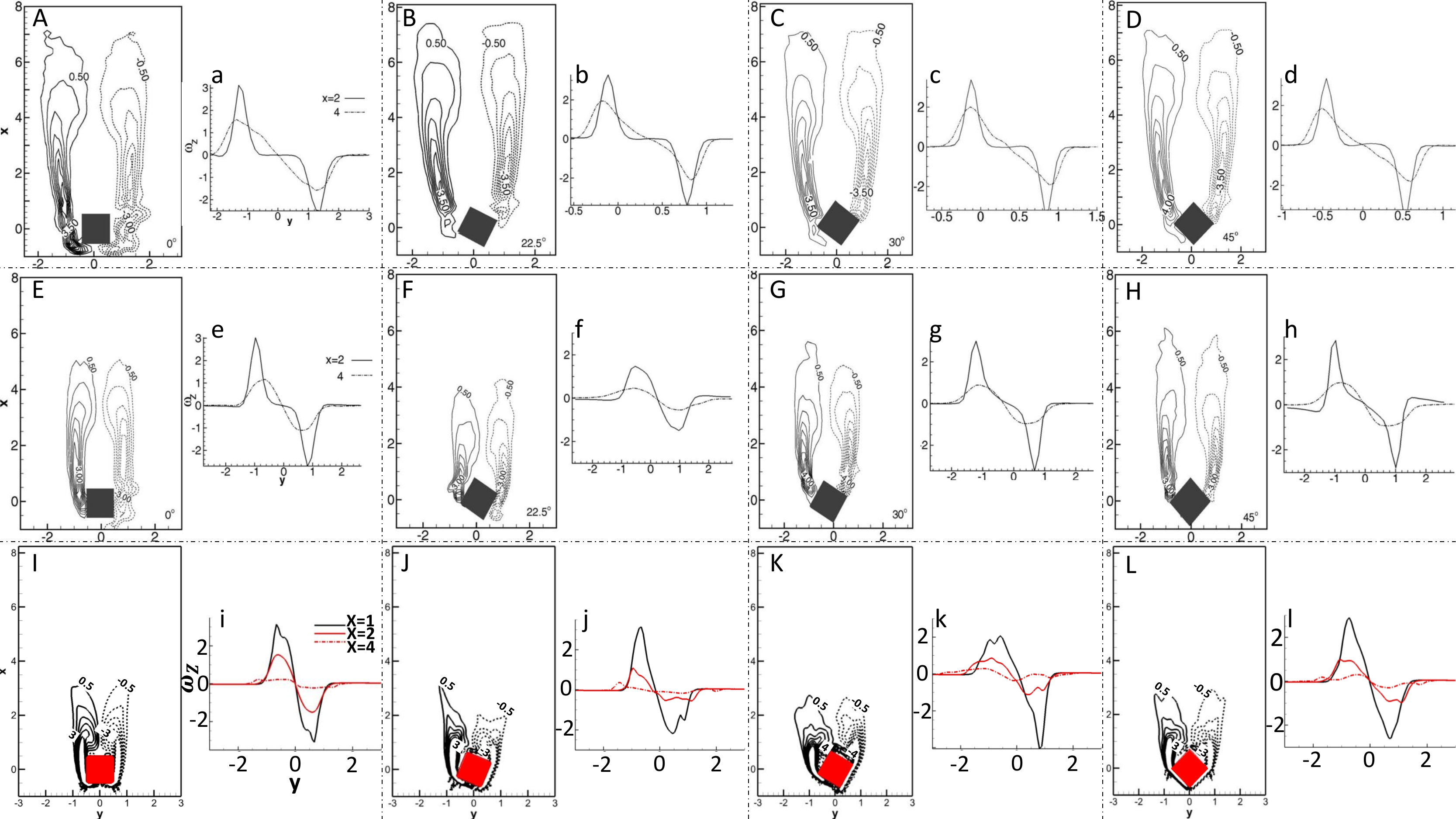}
	\caption{Time-average vorticity ($\omega_z$), capital letters represent the vorticity contour, while lower-case letters represent the vorticity profile. Pictures A-D and a-d show the vorticity for four cylinder orientations with AR = 16. Pictures E-H and e-h show the vorticity for four cylinder orientations with AR = 28. Pictures I-L and i-l show the numerical vorticity for four cylinder orientations in our simulation where AR = $\infty$. In the vorticity contours, solid lines show positive vorticity while dashed lines represent negative vorticity, and $\Delta \omega_z$ = 0.5.}
	\label{fig:vorticity}
\end{figure}

Figure \ref{fig:centerline} compares the centerline recovery of the streamwise velocity and decay of transverse velocity for experiments and numerical simulation with various cylinder orientations. We plot the nondimensional streamwise velocity, $u_{\text{CL}}/u_\text{inf}$ along the x axis at the centerline and the nondimensional transverse velocity, $v/u_\text{inf}$, along the x axis at a offset location from the centerline (y=1). In agreement with the experiments, the time-average streamwise velocity in our simulations is zero on the square and is negative in the recirculation zone. Later, it increases and reaches an asymptotic value, which is in the range of 0.6-0.65 for all aspect ratios and all cylinder orientations. We obtain similar agreement between experimental and numerical results for the transverse velocity that switches from the maximum positive value to a negative value in the recirculation zone and later decays to approximately zero. Comparing the velocity profiles between the cases with different aspect ratios, we observe that both centerline recovery and the decay of transverse velocity are faster at the higher aspect ratio. The asymptotic limit of u$_{\text{CL}}$ and the zero value of v are reached at around x = 15, 7, and 5, respectively for AR = 16, 28 and $\infty$. This trend is in accordance with the larger recirculation vortex at the lower aspect ratio, as shown in Figure \ref{fig:streamline}.

Figure \ref{fig:vorticity} shows the contours (represented by capital letters) and the profiles (represented by lower-case letters) of the time-average nondimensional vorticity for three aspect ratios and four cylinder orientations. In the numerical simulation, we compute the 2D nondimensional vorticity through direct differentiation of the velocity field,
\begin{eqnarray}
	{(\omega_z)}_{i,j} = \frac{(v_{i+1/2,j}-v_{i-1/2,j})/u_{\text{inf}}}{\Delta x/D} - \frac{(u_{i,j+1/2}-u_{i,j-1/2})/u_{\text{inf}}}{\Delta y/D} \ ,
\end{eqnarray} 
where $u$ and $v$ represent the streamwise velocity and the traverse velocity, respectively. In agreement with the experimental record, the maximum value of the numerical vorticity is observed at the cylinder corner (see pictures I-L). A comparison between the numerical and experimental vorticity profiles demonstrates that the vorticity magnitudes in the immediate vicinity of the cylinder are quite similar for different aspect ratios. Consistent with the observation on the recirculation vortex in Figure \ref{fig:streamline}, the smaller spreading of the vortices occurs at higher aspect ratios, and the vortices are asymmetric when the cylinder orientation is 22.5 or $30^\circ$.

\begin{figure}[h!]
	\centering
	\includegraphics[width=0.8\textwidth]{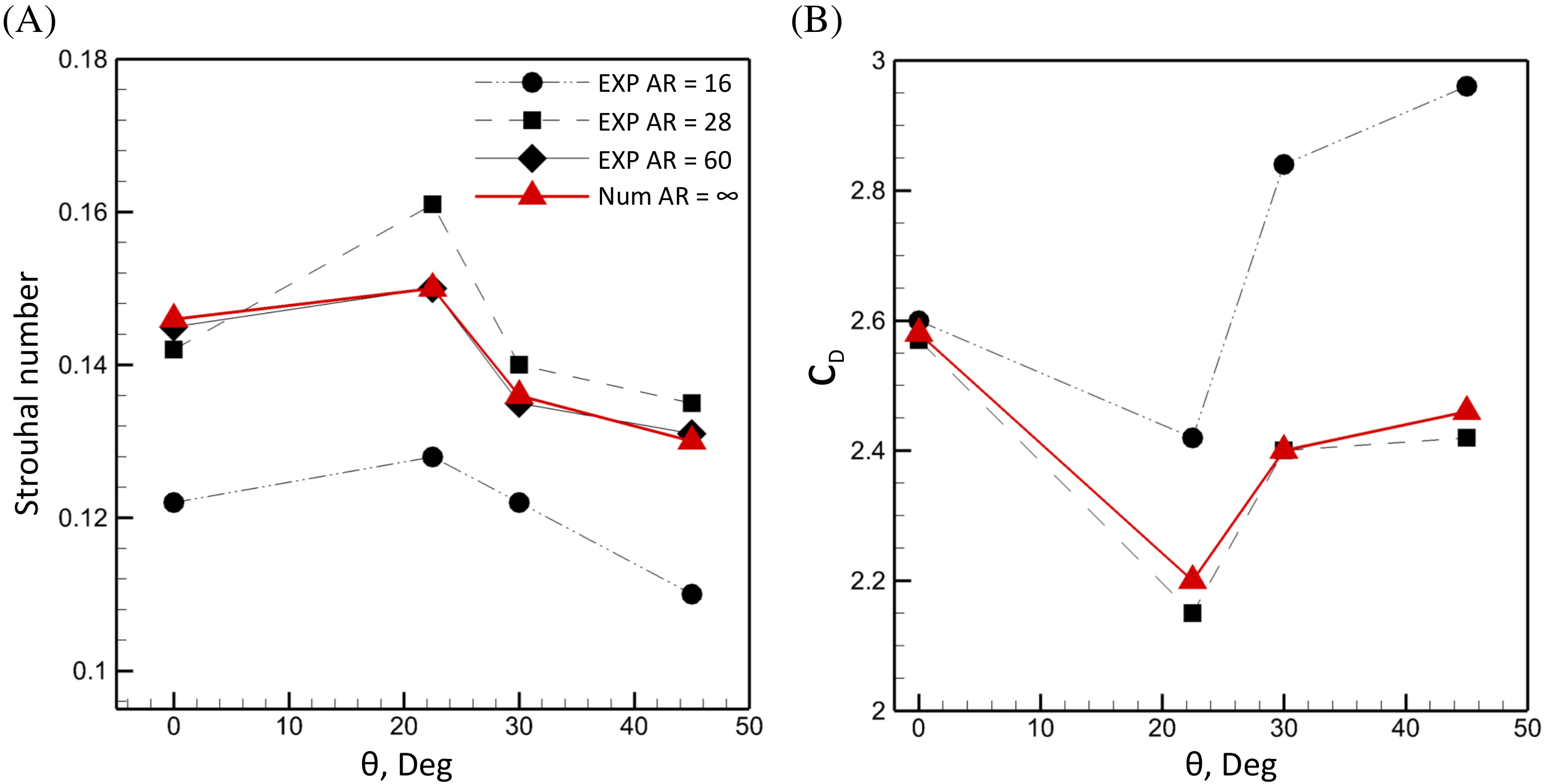}
	\caption{(A) Comparison of time-averaged Strouhal number among aspect ratios (AR) at $\mathrm{Re} = 410$; (B) Comparison of time-averaged drag coefficient among aspect ratios at $\mathrm{Re} = 410$. Both Strouhal number and drag coefficient vary with cylinder orientation. The case labeled as AR=$\infty$ represents the 2D simulation. Dutta et al. \citep{Dutta2009} provided the experimental measurement of Strouhal number for AR = 16, 28, and 60 while only provided the experimental drag coefficient for AR = 16 and 28. }
	\label{fig:drag}
\end{figure}

Figure \ref{fig:drag}A compares the numerical Strouhal number to the experimental values. We find good agreement between the numerical and experimental Strouhal number at the highest aspect ratio available experimentally, $\mathrm{AR} = 60$. The numerical simulation correctly reproduces the qualitative trend, observed in experiments for all aspect ratios, that the Strouhal number increases from the $0^\circ$ cylinder orientation angle to $22.5^\circ$ but drops subsequently with a further increasing angle. The maximum Strouhal number arises around $22.5^\circ$. Dutta et al., \citep{Dutta2009} explained the angular variations in the Strouhal number through an angular dependence in the projected dimension of the cylinder with respect to the incoming flow.

Figure \ref{fig:drag}B shows the variation of the time-average drag coefficient with respect to the cylinder orientations. In 2D numerical simulations, the hydrodynamic drag force, $F_D$, acting on the cylinder cross-section in streamwise direction, is given by
\begin{eqnarray}\label{eq:Bech7}
	F_D=\int_S (-p + 2\mu_fu_x)n_x+\mu_f(u_y+v_x)n_ydS \ ,   
\end{eqnarray}
where $S$ represents the circumference of the cylinder cross-section. Hence, we compute the drag coefficients as,
\begin{eqnarray}\label{eq:Bech9}
	C_D=\frac{2F_D}{\rho_fu^2_{inf}D} \ .   
\end{eqnarray}
In agreement with the experimental results, our simulation shows that the minimum drag coefficient occurs at $22.5^\circ$ angle. Contrary to the Strouhal number, the drag coefficient decreases from $0^\circ$ to $22.5^\circ$ and increases again for larger angles. 

\subsection{Jamming in shear suspensions}\label{sec:CD}
Earlier studies, including numerical \citep{Brady1988} and experimental studies \citep{Karnis1966}, observed that circular particles in a sheared suspension are prone to clustering, which alters the rheology of the suspension significantly. Assuming spherical particles, Brady and Bossis \citep{Brady1988} simulated suspension behavior in a Couette device. They observed that particle clusters at intermediate particle fraction begin to span the gap between the two plates, resulting in a boundary-dominated flow where the suspension moves approximately as a plug with an average speed of approximately half of the imposed shear speed. Karnis et al. \citep{Karnis1966} experimentally observed similar behavior in both shear flow in a cylindrical Couette device and pressure-driven flow in tubes. 
\begin{figure}[h!]
	\centering
	\includegraphics[width=1\textwidth]{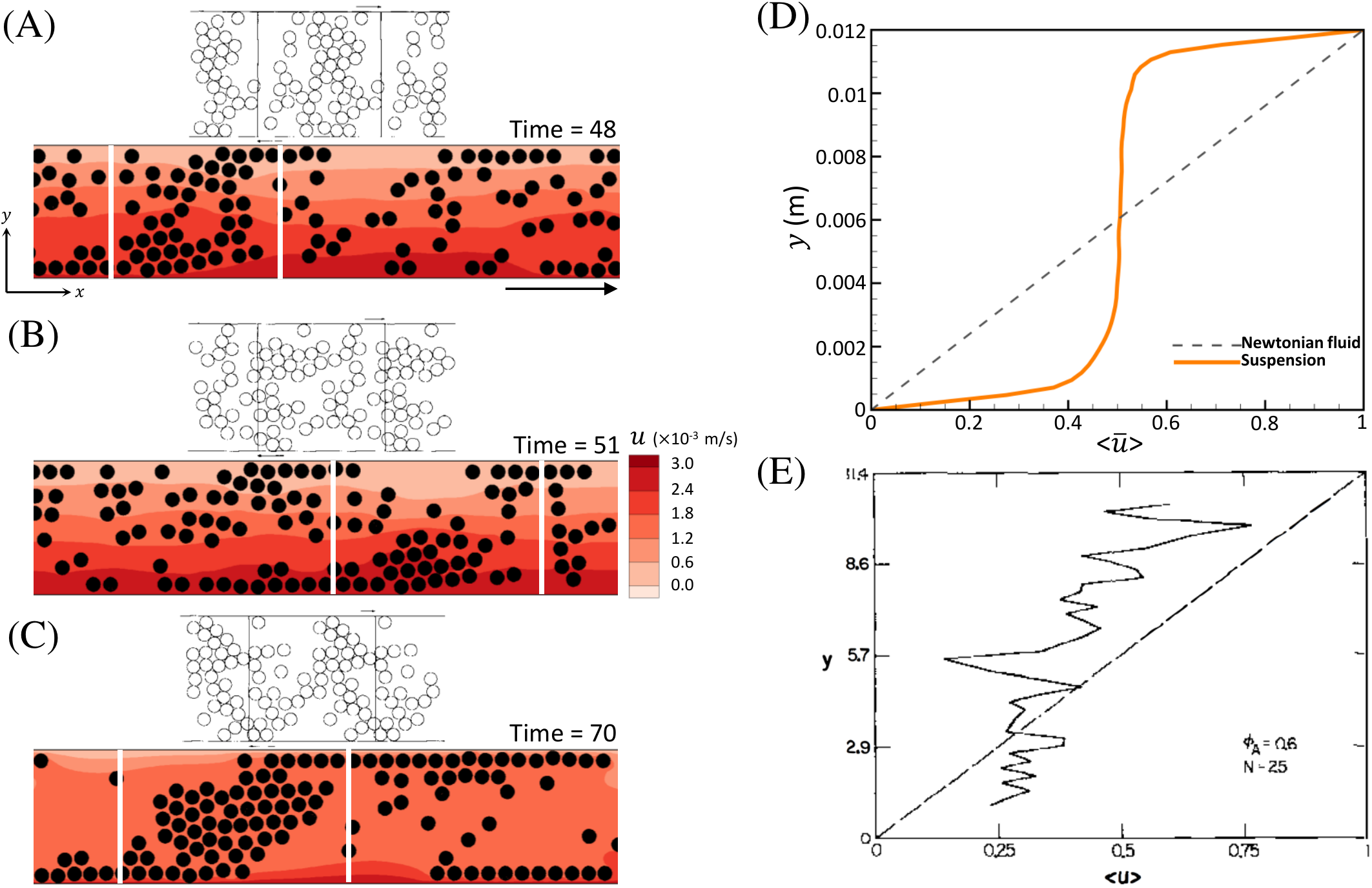}
	\caption{(A)-(C) Snapshots of instantaneous crystal positions for a suspension sheared in a Couette device. The simulation by Brady and Bossis \citep{Brady1988} are reproduced on top. In our results, the color scale represents the horizontal component of the velocity. The time has been nondimensionalized by the shear rate $U/L$, where $U$ is the velocity of the moving plane and $L$ is the width of the gap. (D) Plot of the average velocity in the flow direction versus the transverse coordinate $l$. (E) Profile of the average particle velocity in the flow direction versus the transverse coordinate $l$, shown in Figure 15 of \citep{Brady1988}.}
	\label{fig:plug_behavior_like}
\end{figure}

In our numerical model, we use two elongated, rectangular particles to represents the parallel planes of a Couette cell. We fix the upper particle, $\alpha$, and move the lower particle, $\beta$, with a constant speed $U$. We set the width of the gap between plates, $L$, to $L$ = 0.012 m. Since the motivation for our numerical method is to simulate crystal-bearing lava flows, our parameters are inspired by a magmatic context with a fluid viscosity of $\mu_f=1000$ Pa s and a density of $\rho_f=3000$ kg/m$^3$. With these material properties, the Reynolds number is approximately $\sim 10^{-5}$. We then introduce 120  neutrally buoyant, circular particles with radii of $r=1/20L$. 

Figure \ref{fig:plug_behavior_like} (A)-(C) show three snapshots in time of the instantaneous crystal configurations at a solid area fraction of $\psi\approx0.20$. The color scale represents the velocity in the flow direction. As a comparison, three crystal configuration are reproduced from Brady \citep{Brady1988}. Initially, the particles are uniformly distributed in the domain. Soon, they start forming clusters spanning from the lower to the upper plate (Figure \ref{fig:plug_behavior_like}A). The number of particles in the cluster increases with time (Figure \ref{fig:plug_behavior_like}B) until most of the particles are clustered (Figure \ref{fig:plug_behavior_like}C). At this point, the particles translate more or less as a single entity or plug. We plot the average nondimensional velocity, $<\bar{u}>$, of the suspension in the flow direction versus the transverse coordinate $y$ in Figure \ref{fig:plug_behavior_like} (D). The result by Brady \citep{Brady1988} is shown in (E).
\begin{figure}[h!]
	\centering
	\includegraphics[width=0.9\textwidth]{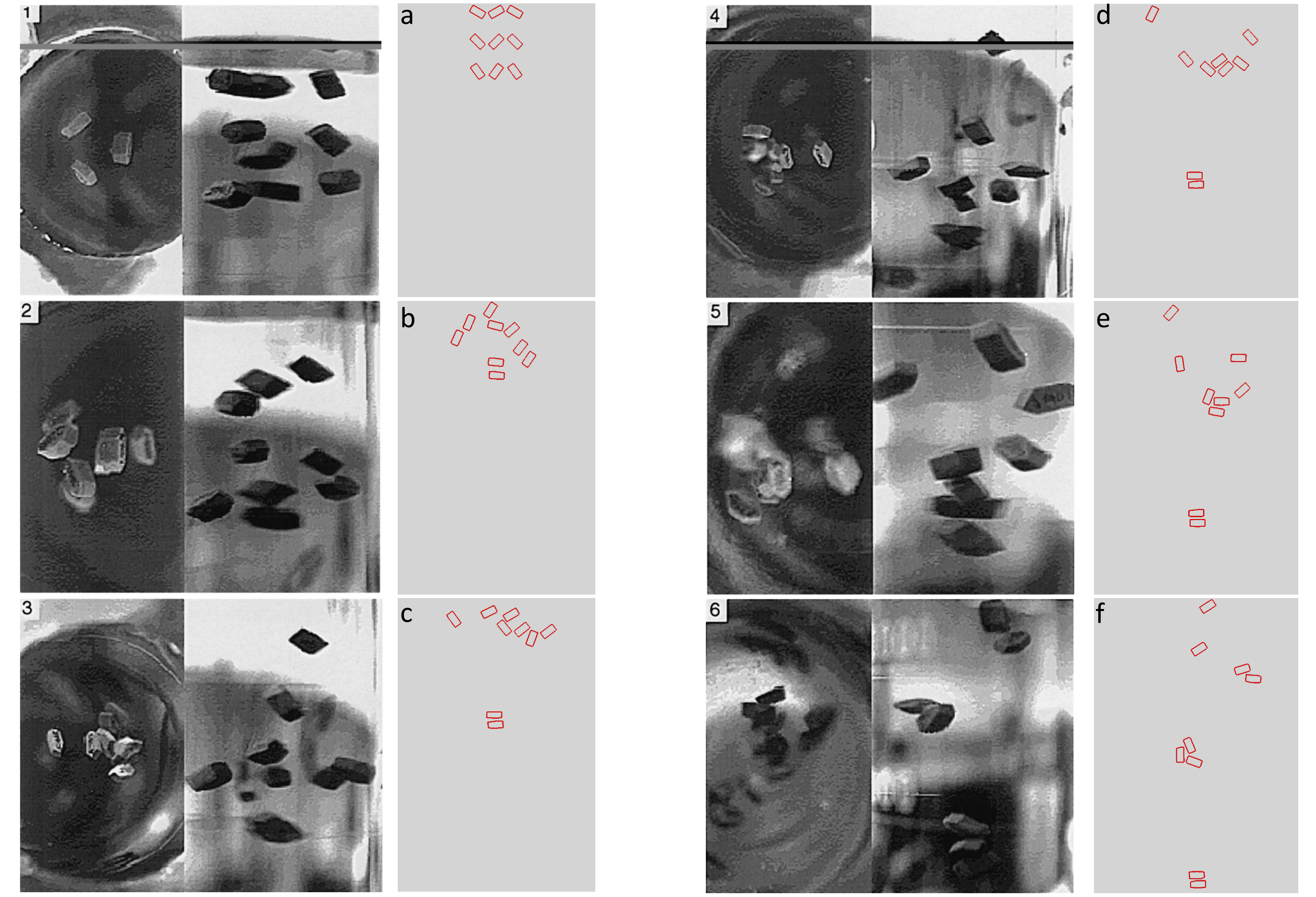}
	\caption{A dilute suspension of nine particles drops in three rows. Pictures 1-6 reprsents the analogue experiment, the left photo of the six pair is a view from on top, the right is aview from the side. Pictures a-f represent the numerical simulation. The photos and the corresponding snapshots are taken at 15-min intervals. All particles have the same density, 1.423 g/cm$^3$.}
	\label{fig:schwindinger_9}
\end{figure}

Following Brady \citep{Brady1988}, we define $<\bar{u}>$ as,
\begin{eqnarray}\label{eq:PFLB1}
<\bar{u}> = \frac{\bar{u}-U}{0-U} \ ,
\end{eqnarray}
where $\bar{u}$ represents the average dimensional velocity in the flow direction, $U$ is the speed of the lower plane and 0 is the speed of the upper plane. For comparison, we plot  $<\bar{u}>$ for the pure fluid flow as a dashed line in (D). We obtain good agreement with \citep{Brady1988} and \citep{Karnis1966} in reproducing that the suspension, including the crystal and the fluid, moves at roughly half the imposed boundary speed in the plug-flow limit. As a consequence, two regions of rapid shear form in the immediate vicinity of the bounding planes.

\subsection{Settling and aggregation of rectangular particles}\label{sec:Sch}
We validate the ability of our numerical method to capture the dynamic interactions between moving rectangular particles by reproducing the laboratory experiments by Schwindinger \citep{Schwindinger1999}. In this suite of experiments, prism-shaped particles are dropped into cold Karo syrup to observe settling behavior. Our simulations use the experimental parameters with crystal sizes of $l_1\times l_2=0.5 \times 0.25$ cm, crystal densities of $\rho_p$=1.372-1.423 ($\pm$0.004) g/cm$^3$, a constant fluid density of $\rho_f = 1.37$ g/cm$^3$, and a viscosity of $\mu_f = 35$ Pa s. The simulations are carried out in a 2D rectangular computational domain which has the same dimeter as the cylinder used in the experiments. We apply a no-slip boundary condition to the wall of the computational domain. In the simulation and the analogue experiments, the particle Reynolds number is $\sim$ 10$^{-6}$.
\begin{figure}[h!]
	\centering
	\includegraphics[width=1.0\textwidth]{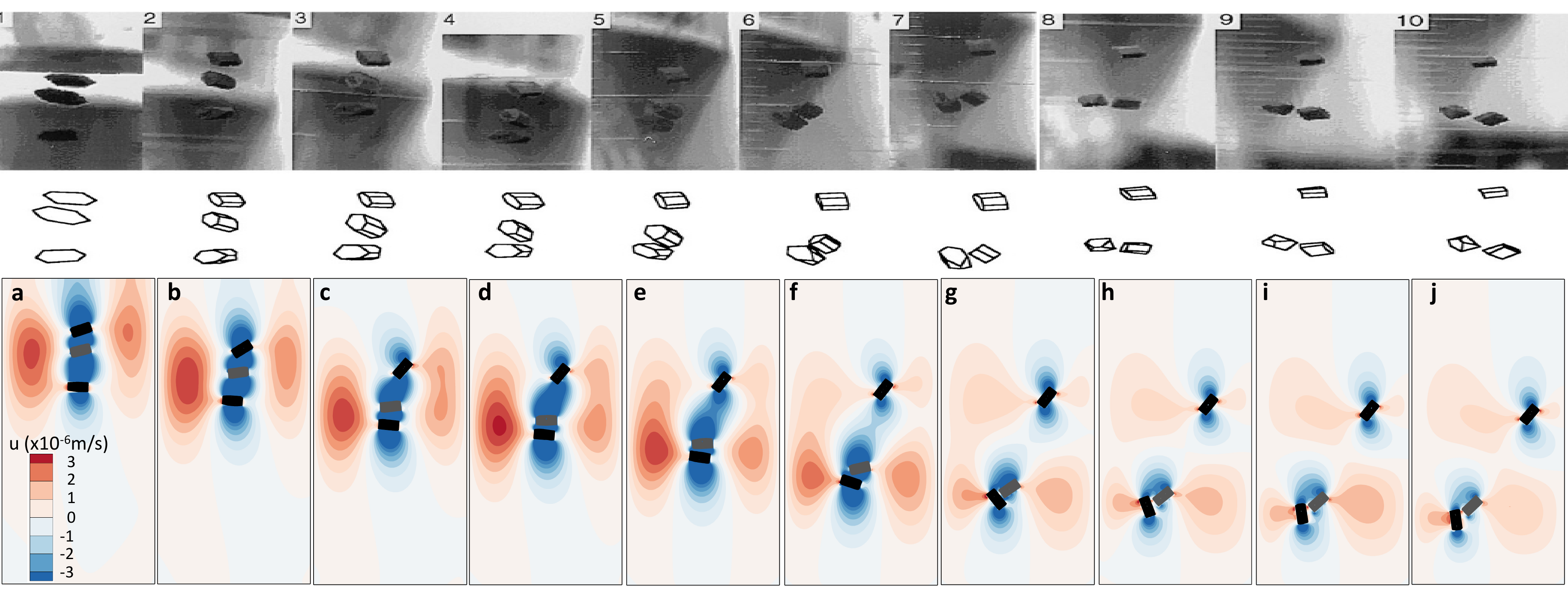}
	\caption{Experimental run and its 2D numerical reproduction in which three particles drops vertically. The photos and the corresponding snapshots are taken at 15-min intervals. The leading and tailing particles are denser than the middle particle. The leading and the trailing particles are represented by black rectangular, while the middle one is silver.}
	\label{fig:schwindinger_10}
\end{figure}

Figure~\ref{fig:schwindinger_9} 1-6 shows an experiment recording the settling-induced interactions between nine particles. Initially, the particles are arranged in three rows (Figure~\ref{fig:schwindinger_9} 1a). Shortly after the onset of motion, they aggregate into clusters (Figure~\ref{fig:schwindinger_9} 1b). One of the crystal clusters separates from the larger aggregate (Figure~\ref{fig:schwindinger_9} 1c, d) and settles more rapidly. Later, other clusters of two to three particles separate from the other particles and settle (Figure~\ref{fig:schwindinger_9} 1e,f). Different from spheres, which tend to form polygonal arrangements \citep{Schwindinger1999}, the non-spherical particles are drawn into the center of the domain and create pairs or triplets. Our numerical simulations show similar behavior as the experiments. 

\section{Particle rotation may alter suspension behavior} \label{sec:AP}

\begin{figure}[h!]
	\centering
	\includegraphics[width=0.9\textwidth]{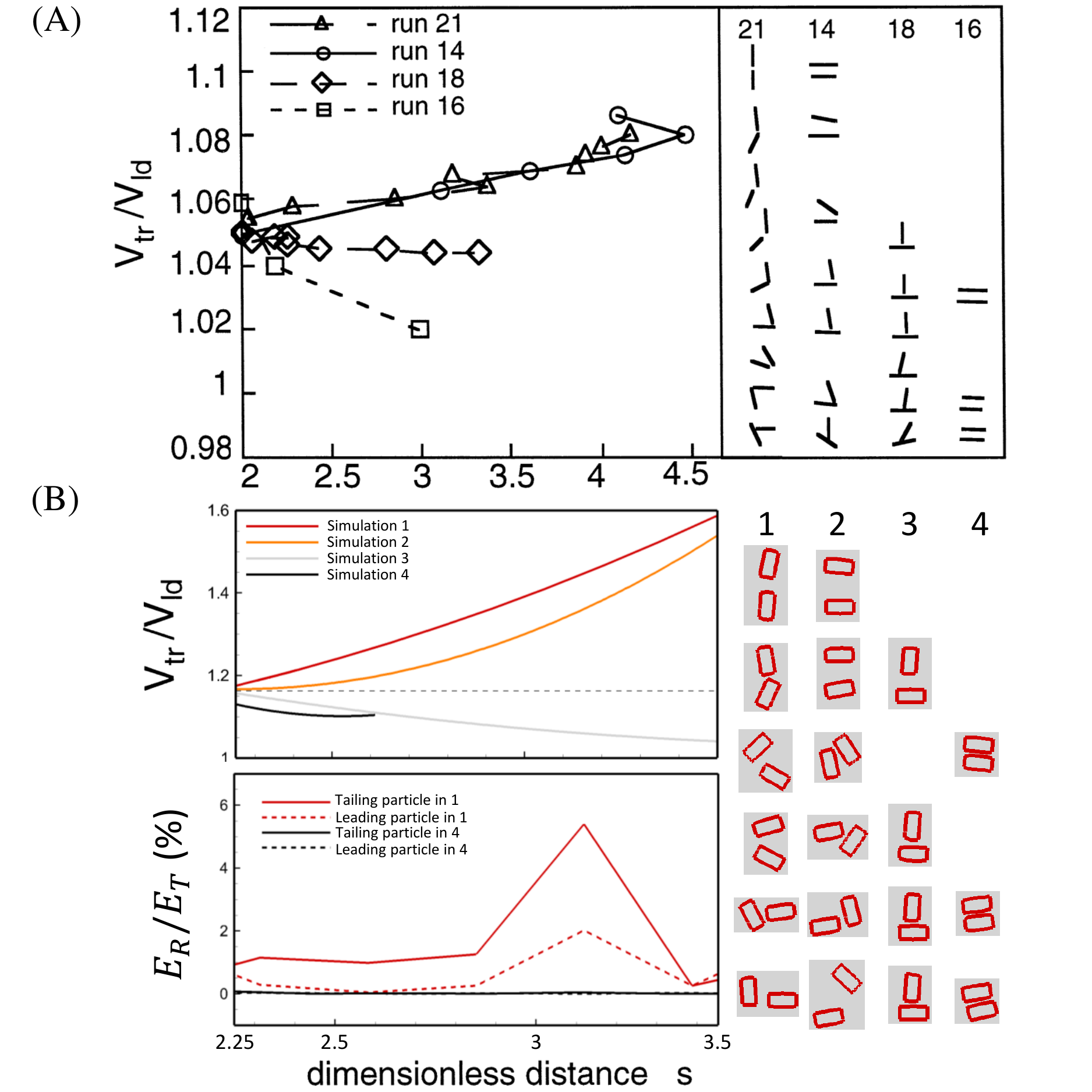}
	\caption{Relationship between the velocity ratio and the behavior of two colliding rectangular particles as seen in experiments (A) and simulations (B). (A) Experiments by Schwindinger \citep{Schwindinger1999} showed that the ratio of settling speed for the tailing, $V_{tr}$, as compared to the leading particle, $V_{ld}$, increases with separation distance when the particles are tumbling (Runs 14 and 21) and decreases with separation distance when the particles are not tumbling (Runs 16 and 18). (B) In simulations 1, 3 and 4, we use $\rho$ = 1.423 g/cm$^3$ to both particles and initially arrange them to mimic the initial positions in experiments 21, 18 and 19, respectively. In simulation 2, we use a slightly higher density $\rho_{\text{tr}}$ = 1.423 g/cm$^3$ for the trailing particle as compared to the leading particle ($\rho_{\text{ld}}$ = 1.403 g/cm$^3$) and use the initial configuration from experimental 14. The separation distance, s, is nondimensionlized by the width of the particles.}
	\label{fig:collision_velocity}
\end{figure}

The behavior observed in the Schwindinger experiments \citep{Schwindinger1999} is reminiscent of drafting, kissing, and tumbling \citep{fortes1987nonlinear}, where the low-pressure wake behind a leading particle drafts in a trailing particle of comparable density until the two came in contact and then tumble apart. At very low Reynolds ($\mathrm{Re} \ll 1$), however, this mechanism no longer applies, because of the absence of low-pressure wakes. Nonetheless, Schwindinger observed that a tailing, lighter particle can catch up with a leading, denser particle even at approximately zero Reynolds number. Figure \ref{fig:schwindinger_10} 1-10 shows an experiment where three particles is dropped into the Karo syrup. In this particular experimental setup, the tailing and leading particles ($\rho$ = 1.423 g/cm$^3$) are slightly denser than the middle particle ($\rho$ = 1.372 g/cm$^3$). Schwindinger reports similar behavior even for equal densities.

In agreement with the analogue experiment, our simulations in Figure~\ref{fig:schwindinger_10} a-j show that the middle particle accelerates through the meniscus of seemingly disturbed fluid between the tailing and leading particles and catches up with the leading one. The reason for this behavior, however, is not related to spatial pressure variations but to the long-range interactions of the particles. As shown in Figure \ref{fig:schwindinger_10} a-j, the middle particle experiences a mean flow field that is already downward oriented. Its motion is a superposition of mean-field transport and individual settling and hence faster than buoyancy-driven settling alone. 

An indirect indication of this mechanism is the differing rotational behavior observed by Schwindinger \citep{Schwindinger1999}. Figure \ref{fig:collision_velocity} A shows the observed ratio of settling speeds for the tailing, $V_{tr}$, and the leading, $V_{ld}$ particle. In all of the shown experiments, the tailing particles approach the leading particles ($V_{tr}/V_{ld}>1$), but at different rates. High approach rates (i.e., experiments 21 and 14) are associated with more pronounced rotation of the particles as visually evident in the plotted interface positions and quantified more systematically in Figure \ref{fig:collision_velocity} B. We compute the rotational energy by $E_r=1/2I_p\omega_p^2$ and compute the translational energy by $E_T=1/2M_pV_p^2$, where $I_p$ is particle's angular moment of inertia tensor, $\omega_p$ is its angular velocity, $M_p$ is its particle mass and $V_p$ is its velocity. Rotations are indicative of unsteadiness in the mean flow since isolated  non-spherical particles do not rotate when immersed in a steady-state flow at zero Reynolds number \citep{becker1959effects}. Rotations are hence suggestive of variability in the mean flow field, which in this particular case, leads to a relative speed-up of the tailing particle.

\begin{figure}[h!]
	\centering
	\includegraphics[width=1\textwidth]{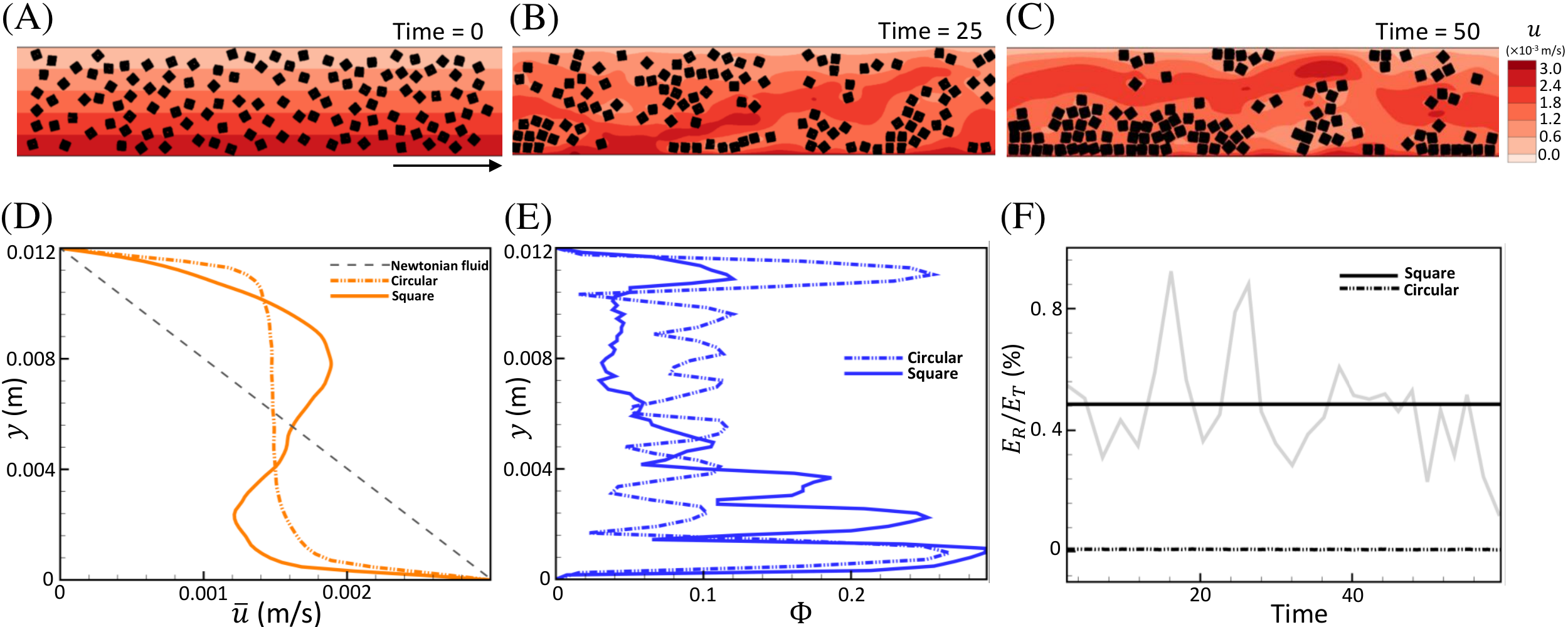}
	\caption{(A)-(C) Snapshots of instantaneous square particle configurations for the sheared suspensions. The color scale indicates the velocity in the flow direction. (D) represents the comparison of velocity profile between the suspension with circular and square crystals. Silver dashed line represents the pure Newtonian fluid; dashed-dot-dot curve represents the circular crystal suspension; solid curve represents the square crystal suspension. (E) represents the comparison of solid fraction between the suspension with circular and square crystals. (F) Ratios of the rotational energy to the translational energy. The time has been nondimensionalized by the shear rate $U/L$, where $U$ is the velocity of the moving plane and $L$ is the width of the gap. Light silver curves represent the datas recorded from the simulation, while black curves represents their mean value.}
	\label{fig:plug_circ_sqrt}
\end{figure}

While rotations are a proxy for existing variability in the mean flow, they also generate variability in the mean flow. This effect is particularly pronounced in experiment 21 and simulation 1 from Figure~\ref{fig:collision_velocity}, where both particles experience significant rotation due to their interactions, and most suppressed in experiment 16 and simulation 4, where the crystals are in close contact from the beginning and aligned in a way that makes them less susceptible to rotation. Contrary to the well-known drafting, kissing, and tumbling behavior at finite Reynolds number \citep{fortes1987nonlinear}, this zero-Reynolds variant of drafting, kissing, and tumbling (e.g., experiments 21 and simulation 1) is hence highly dependent on crystal shape and the initial position of the crystals. Nonetheless, it is another instance of inertia-like behavior in solid-bearing suspensions at zero Reynolds number and highlights the nonlinearity and stochasticity introduced into these flows through long-range particle interactions analogous to the effect of turbulence at high Reynolds number \citep{xue1992diffusion, tong1998analogies, levine1998screened}.

One could argue that in the Schwindinger experiments \citep{Schwindinger1999}, rotation is mostly a diagnostic of spatial variability in the mean flow. It is helpful for understanding the tendency of particles to aggregate and differences in settling speed, but it is not clear whether it is an essential component of the dynamics that a large-scale model of particle settling would need to capture. Potentially, the behavior of the clay particles in \citep{Schwindinger1999} could be approximated by considering an assemblage of spheres with variable size to represent clustering. At low solid fraction, that argument has some merit. It is not clear whether it generalizes to larger solid fractions, however, where jamming becomes important, requiring stable, long-term contact between particles to build force chains.

To test the importance of rotation in this regime, we revisit the simulation of a suspension sheared in a Couette device from Section \ref{sec:CD} using square crystals. In the interest of easier comparability, we keep the overall solid fraction the same ($\psi = 0.2$) and define our square particles such that a single square has approximately the same area as one of circular particles used previously. Figure \ref{fig:plug_circ_sqrt} shows that the plug-like flow observed in Figure~\ref{fig:plug_behavior_like} for circular particles no longer occurs. In fact, the flow field in the fluid phases changes rather profoundly from the linear profile typical of Couette flow (Figure~\ref{fig:plug_behavior_like}A) to a partially channelized flow field with a maximum flow speed in the center of the domain, not unlike Poiseuille flow (Figure~\ref{fig:plug_circ_sqrt}C) if unsteady. 

\begin{figure}[h!]
	\centering
	\includegraphics[width=1\textwidth]{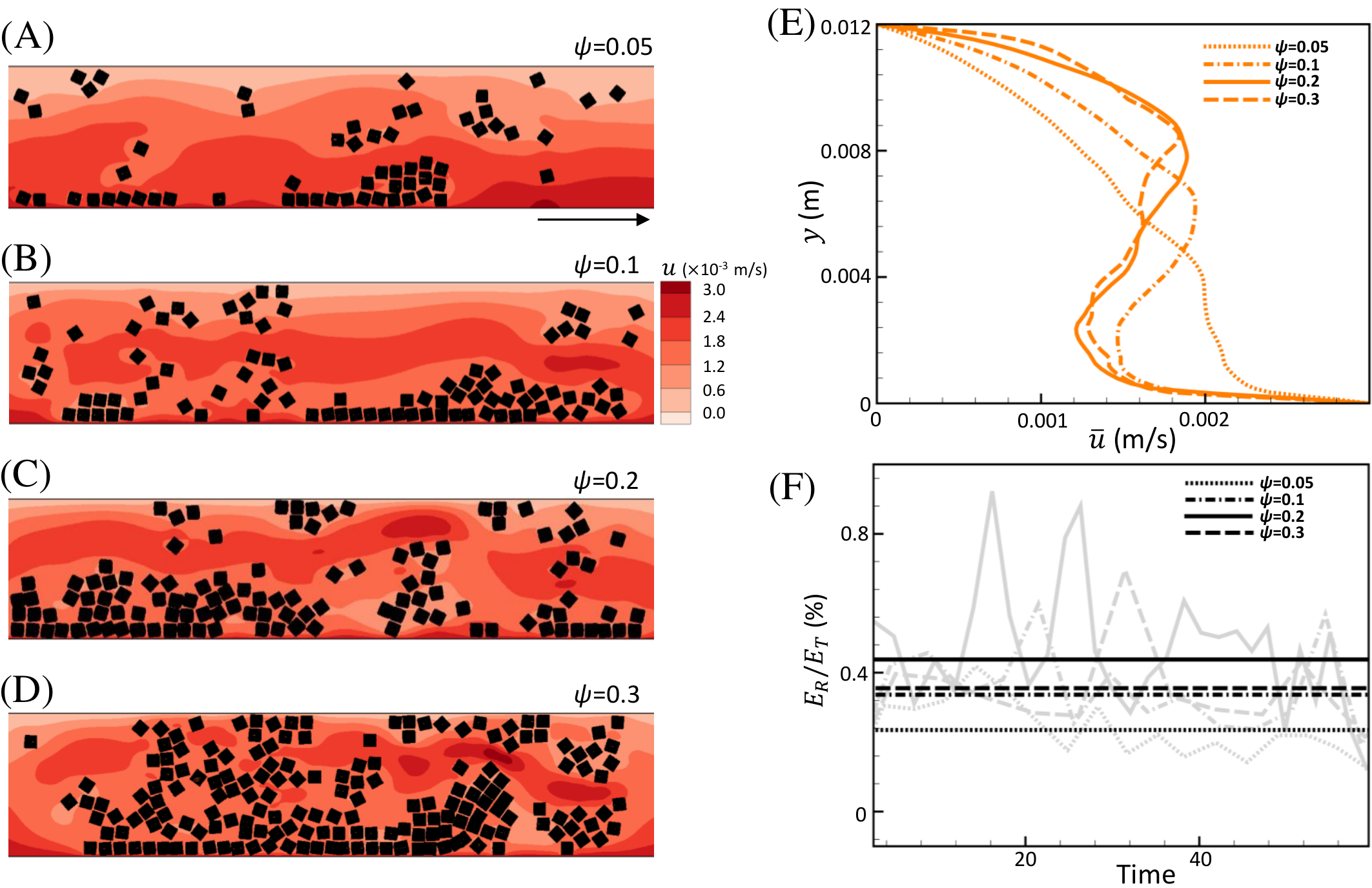}
	\caption{(A)-(D) Snapshots of instantaneous square particle configurations for the sheared suspensions with different solid areal fractions, $\psi$. The color scale indicates the velocity in the flow direction. (E) represents the comparison of the velocity profiles with different solid fractions. Dot line represents the suspension with $\psi$=0.05; dashed dot line for $\psi$=0.1; solid line for $\psi$=0.2; dashed line for $\psi$=0.3. (F) Ratios of the rotational energy to the translational energy with different solid fractions. The time has been nondimensionalized by the shear rate $U/L$, where $U$ is the velocity of the moving plane and $L$ is the width of the gap. Light silver curves represent the datas recorded from the simulation, while black curves represent their mean value.}
	\label{fig:plug_sqrt}
\end{figure}

Figure \ref{fig:plug_circ_sqrt}D compares the profile of the velocity in the flow direction averaged over the transverse direction, $\bar{u}$, for circular and square particles. The velocity profile in the suspension of square crystals has two peaks at approximately $y = 0.002$ m and $y = 0.008$ m. The reason for this behavior is that the square particles do not distributed equally throughout the domain as highlighted in Figure \ref{fig:plug_circ_sqrt}E, where we plot average solid fraction averaged in the flow direction for the two suspensions. The circular crystal suspension reaches maximum solid fraction in the two rapid shear regions in the immediate vicinity of the bounding plates. In contrast, the square particles accumulate predominantly in the lower portion of the domain, thereby leaving space for almost solid-free and hence comparatively rapid flow in the upper third of the domain. In the example shown here, the flow speed reaches its peak value around $y = 0.008$ m in the approximately particle-free channel.

In this case, the particle shape has altered flow behavior more fundamentally than in the Schwindinger experiments \citep{Schwindinger1999}. Aggregation within the solid phase has forced a shift from Couette-type flow to Poiseuille-type flow in the fluid phase. Our simulations suggest that the increased tendency of square particles to rotate is essential for creating this shift. Figure \ref{fig:plug_circ_sqrt}F shows the ratio of rotational to translational energy for square as compared to circular particles. While the ratio of kinetic energy stored in rotations is low as compared to translation, rotations have a strong impact on the dynamics, because they destabilize force chains. As a consequence, force chains are much more long-lived for circular as opposed to square particles. Instead of forming force chains throughout the domain, square particles aggregate and form clusters at the bottom of the domain, aided by their much higher random packing fraction (e.g., Figure~\ref{fig:plug_behavior_like}C). The flow is diverted around the crystal-rich zone and channelizes, which in turn makes it more difficult for a force chain to build across it.

Evidently, this effect is dependent on the solid fraction of the suspension as illustrated in Figure~\ref{fig:plug_sqrt}. At the lowest solid fraction, $\psi = 0.05$, the Couette flow is only slightly disturbed by the motion of the crystals (Figure~\ref{fig:plug_sqrt} A), but starts to collapse to an unsteady Poiseuille-type profile at $\psi = 0.1$ as cluster form at the two bounding plates. For $0.1 \le \psi \le 0.2$, the Poiseuille flow profile occupies most of the computational domain, but it increasing localizes around $\psi = 0.3$. At this stage, several large particle clusters form that tend to break up the flow in some locations. Nonetheless, a Poiseuille-type flow persists in spatially and temporally averaged sense even at $\psi = 0.3$ (see Figure~\ref{fig:plug_sqrt}D). We compare the ratio of rotational to translational energy for the particles at different solid areal fractions to find that rotations remain pronounced until $\psi = 0.3$ (Figure~\ref{fig:plug_sqrt}F). The rotational to translational energy ratio is lowest at $\psi$=0.05 because particles are, on average, separated the furthest in that case and interactions are less pronounced. 

\section {Conclusion}
The goal of this study is to better understand how rotations of non-spherical particles alter suspension behavior. Since the stochasticity of solid-bearing suspensions emerges from the long-range interactions between particles, we develop a direct numerical technique for fully resolving solid-fluid coupling at the scale of individual interfaces. The broader motivation for this work is to better understand and model hazardous, crystal-bearing lava flows and we hence focus specifically on rectangular and square particles commonly found in field samples. Our numerical method relies on distributed Lagrange Multipliers to enforce rigid-body motion of the particles, coupled with an immersed interface method to correctly enforce the no-slip constraint on the solid interfaces. To prevent overlap between particles, we build a collision model that applies to both circular and rectangular shapes and correctly computes the translational and rotational momentum transfer during collision. We have verified and validated the efficiency and accuracy of our method at low and intermediate Reynolds number. Our simulations show that particle interactions tend to amplify rotational motion of square and rectangular crystals both at low and intermediate solid fraction. The consequences range from introducing unexpected dynamics such as drafting, kissing and tumbling behavior at zero Reynolds number to  completely altering the flow field in the fluid. While the current model is 2D, these effects are likely to persist in 3D, since the rotational energy of non-spherical particles would tend to be greater and hence potentially even more important dynamically in higher dimensions. 
 
\section*{Acknowledgement}
This research was supported by the US Army Research Office through grant W911NF-18-1-0092. We would like to thanks Paul Wallace for encouraging us to look at crystal synneusis and Katherine V. Cashman for permission to use her thin-section images in Figure 1.

\section*{Author contributions}
Z. Q. developed the code, performed the numerical simulations and created the figures. K. A. developed the collision model for rectangular particles. J. S. contributed to code development as well as study conception and analysis. All authors contributed to the text.

\vspace{20pt}
\noindent

\section*{Reference}

\end{document}